\documentclass[aps,prd,twocolumn,
preprintnumbers,superscriptaddress,
showpacs,floatfix]{revtex4}

\usepackage{bm}
\usepackage{epsfig}
\usepackage{graphics}

\newcommand{\eeq}{\end{equation}}
\newcommand{\beq}{\begin{equation}}
\newcommand{\ba}{\begin{array}}
\newcommand{\ea}{\end{array}}
\newcommand{\bea}{\begin{eqnarray}}
\newcommand{\eea}{\end{eqnarray}}

\newcommand{\eps}{\epsilon}

\newcommand{\lsim}{\mbox{\raisebox{-.9ex}
{~$\stackrel{\mbox{$<$}}{\sim}$~}}}
\newcommand{\gsim}{\mbox{\raisebox{-.9ex}
{~$\stackrel{\mbox{$>$}}{\sim}$~}}}

\begin{document}

\preprint{UT-STPD-3/05}

\title{Modular inflation and the orthogonal
axion as curvaton}

\author{K. Dimopoulos}
\email{k.dimopoulos1@lancaster.ac.uk}
\affiliation{Physics Department, Lancaster
University, Lancaster LA1 4YB, U.K.}
\author{G. Lazarides}
\email{lazaride@eng.auth.gr}
\affiliation{Physics Division, School
of Technology, Aristotle University of
Thessaloniki, Thessaloniki 54124, Greece}

\date{\today}

\begin{abstract}
\noindent
We study a particular supersymmetric realization
of the Peccei-Quinn symmetry which provides
a suitable candidate for the curvaton field.
The class of models considered also solves the
$\mu$ problem, while generating the
Peccei-Quinn scale dynamically. The curvaton
candidate is a pseudo Nambu-Goldstone boson
corresponding to an angular degree of freedom
orthogonal to the axion field. Its order
parameter increases substantially following a
phase transition during inflation. This results
in a drastic amplification of the curvaton
perturbations. Consequently, the mechanism is
able to accommodate low-scale inflation with
Hubble parameter at the TeV scale. Hence, we
investigate modular inflation using a string
axion field as the inflaton with inflation
scale determined by gravity mediated soft
supersymmetry breaking. We find that modular
inflation with the orthogonal axion as curvaton
can indeed account for the observations for
natural values of the parameters.
\end{abstract}

\pacs{98.80.Cq}
\maketitle

\section{Introduction}
\label{sec:intro}

\par
The last decade experienced a plethora of
precise cosmological observations. These
observations have confirmed the basic
predictions of inflation rendering the latter
an essential extension of the hot big bang
model. Precision cosmology upgraded inflation
model-building at the next level, that of
the design of complex realistic models
(in contrast to early single-field
fine-tuned inflationary models). These
models utilize and reflect the rich content of
particle theory. A first such attempt was the
well known hybrid inflation model
\cite{hybrid,cllsw,susyhybrid}, which couples
the inflaton field with the Higgs field of a
grand unified theory (GUT). In this way,
hybrid inflation dispenses with a number of
tuning problems, which plagued most simplistic
single-field inflation models.

\par
In the same spirit, curvaton models
\cite{enqv,LW,moroi} (see also \cite{earlier})
use a second field to generate curvature
perturbations. This curvaton field is not an
{\it ad hoc} degree of freedom introduced by
hand, but, instead, it is \cite{curv,pngb,chun}
a realistic field already present in simple
extensions of the standard model (SM). In
the context of the curvaton, inflation
model-building is substantially liberated
\cite{liber,liber2} allowing for more
realistic and less fine-tuned models.

\par
A particular advantage of inflation
model-building in the context of the curvaton
mechanism is that it is possible to construct
\cite{liber,amplif} models with inflationary
energy scale much lower than the GUT scale.
In this spirit, we investigate, in this paper,
the possibility of using a string axion field
as the inflaton.

\par
The nature and origin of the inflaton field are
still open questions in inflation
model-building. Typically, what is required is
a light field with suppressed
interactions with other fields including those
of the SM. By ``light'', we mean a field whose
effective mass is smaller than the Hubble
parameter $H_*$ at the time when the
cosmological scales exit the horizon during
inflation. This guarantees that inflation lasts
long enough to encompass the cosmological
scales. The interactions of this field have to
be suppressed in order not to lift the
flatness of the scalar potential along the
inflationary trajectory.

\par
However, the case of slow-roll inflation
(i.e. the case with inflaton mass $\ll H_*$)
suffers from the fact that, typically,
supergravity (SUGRA) introduces
\cite{cllsw,randall} corrections to the
inflaton mass of order $H_*$ during the
inflationary period. To keep the inflaton mass
under control, one may use as inflaton a pseudo
Nambu-Goldstone boson (PNGB) field, since the
flatness of the potential of such a field is
protected by a global ${\rm U}(1)$ symmetry.
Promising such candidates are \cite{modular}
the string axions, which are the imaginary
parts of string moduli fields with the flatness
of their potential lifted only by (soft)
supersymmetry (SUSY) breaking. This results in
inflaton masses of order $H_*$. Hence, such
modular inflation is of the fast-roll type
\cite{fastroll}. Fast-roll inflation lasts only
a limited number of e-foldings, which, however,
can be enough to solve the horizon and flatness
problems.

\par
The inflationary energy scale, in this model,
is much lower than the GUT scale. As a result,
the perturbations of the inflaton field are not
sufficiently large to account for the
observations. Consequently, a curvaton field is
necessary to provide the observed curvature
perturbation. However, even the curvaton cannot
\cite{lythbound} generically help us to reduce
the inflationary scale to energies low enough
for modular inflation. This is possible only in
certain curvaton models which amplify
\cite{amplif,lett} additionally the curvaton
perturbations. We describe in detail such a
model belonging to a class of SUSY realizations
of the Peccei-Quinn (PQ) symmetry \cite{pq},
which also solves the strong $CP$ and $\mu$
problems. We use as curvaton an angular degree
of freedom orthogonal to the QCD axion.

\par
Our paper is structured as follows. In
Sec.~\ref{sec:modular}, we present a brief
outline of modular inflation. In
Sec.~\ref{sec:amplification}, we analyze
the amplification mechanism for the curvature
perturbations due to a PNGB curvaton with
varying order parameter. In
Sec.~\ref{sec:quest}, we investigate whether it
is possible to employ in the role of such a
PNGB curvaton an angular degree of freedom
orthogonal to the QCD axion (``orthogonal
axion'') in SUSY theories. We show that the
simplest constructions using two PQ superfields
cannot work because the orthogonal axion is not
appropriately light during inflation. In
Sec.~\ref{sec:model}, we construct in detail an
appropriate class of PQ models involving three
SM singlet superfields, only two of which carry
PQ charges. In Sec.~\ref{sec:potential}, we
study in detail the characteristics of the
scalar potential in the above class of curvaton
models. In Sec.~\ref{sec:curvphys}, we focus
on curvaton physics and derive a number of
important constraints necessary for the model
to be a successful curvaton model. In
Sec.~\ref{sec:ex}, we quantify our findings
in a concrete example of this class of models.
We find that this concrete model can indeed
work for natural values of the model
parameters. Finally, in Sec.~\ref{sec:concl},
we discuss our results and present our
conclusions. Throughout the paper, we use
natural units, where \mbox{$c=\hbar=1$} and
the Newton's gravitational constant is
\mbox{$G=8\pi M_{\rm P}^{-2}$} with
\mbox{$M_{\rm P}\simeq 2.44\times 10^{18}~
{\rm GeV}$} being the reduced Planck mass.

\section{Modular inflation}
\label{sec:modular}

\par
String theory, in general, contains a number
of moduli fields $\Phi_i$, whose tree-level
K\"{a}hler potential, in 4-dimensional
effective SUGRA, is of the form
\begin{equation}
K=-M_{\rm P}^2\sum_i
\ln[(\Phi_i+\Phi_i^*)/M_{\rm P}].
\end{equation}
Hence, the K\"{a}hler potential is flat in the
directions of the imaginary parts of the moduli
$\mbox{Im}(\Phi_i)$. The same is true for the
F-term scalar potential $V_F$, despite the
fact that the superpotential may receive
non-perturbative contributions (e.g. from
gaugino condensation) of the form
\beq
\Delta W\,(\Phi_i)=
\Lambda^3\exp(-\beta_i\Phi_i/M_{\rm P}),
\label{eq:DW}
\eeq
since $V_F$ turns out to be independent of the
phase of $\Delta W$. The mass parameter
$\Lambda$ in Eq.~(\ref{eq:DW}) is usually taken
to be the string scale and the $\beta_i$'s are
model-dependent coefficients of order unity.
The $\mbox{Im}(\Phi_i)$ fields are periodic
(by modular invariance)
\begin{equation}
\mbox{Im}(\Phi_i)\equiv\mbox{Im}(\Phi_i)+
2\pi f_i,
\end{equation}
where $f_i\sim M_{\rm P}$. This is why the
$\mbox{Im}(\Phi_i)$ fields are also called
string axions.

\par
In compactified heterotic string theory, these
axions correspond \cite{Bfield}, in fact, to
the massless modes of the second-rank
antisymmetric tensor field $B$:
\beq
B= b_{\mu\nu}dx^\mu\wedge dx^\nu+
b_I\omega^I_{\alpha\beta^*}dy^\alpha\wedge
dy^{\beta *},
\eeq
where $x^\mu$ are the coordinates in the usual
4-dimensional space, $y^\alpha$ the complex
coordinates in the compactified
extra-dimensional space and
$\omega^I_{\alpha\beta^*}$ harmonic (1,1) forms
parametrizing the geometry of the compactified
space \cite{footnote1}. The usual 4-dimensional
components $b_{\mu\nu}$ of $B$ correspond to
the so-called model-independent axion, while
the extra-dimensional components $b_I$ of $B$
correspond to the so-called model-dependent
axions.

\par
The flatness of the string axion potential is
lifted only by soft SUSY breaking, which tilts
the vacuum manifold by an amount determined by
the SUSY-breaking scale. One can consider
\cite{modular} that it is this potential that
provides the vacuum energy density necessary
for inflation. Assuming gravity mediated soft
SUSY breaking, the inflationary potential $V_*$
at the time when the cosmological scales exit
the horizon is of intermediate scale:
\begin{equation}
V_*^{\frac{1}{4}}\sim \sqrt{m_{3/2}M_{\rm P}}
\sim 10^{10.5}~{\rm GeV}
\end{equation}
for which \mbox{$H_*\sim m_{3/2}$}, where
\mbox{$m_{3/2}\sim 1~{\rm TeV}$} stands for the
gravitino mass. The inflationary potential is
of the form
\begin{equation}
V(s)=V_{\rm m}-\frac{1}{2}m_s^2s^2+\cdots,
\label{Vinf}
\end{equation}
where the ellipsis denotes terms which are
expected to stabilize the potential at
\mbox{$s\sim M_{\rm P}$} with $s$ being the
canonically normalized string axion. Therefore,
in the above formula, we have
\beq
V_{\rm m}\sim (m_{3/2}M_{\rm P})^2\quad
{\rm and}\quad m_s\sim m_{3/2}.
\label{V0}
\eeq
This inflation model results in fast-roll
inflation, where
\bea
&s=s_{\rm i}\exp(F_s\Delta N)\quad{\rm with}\quad
F_s\equiv\frac{3}{2}\left(\sqrt{1+4c/9}-1\right),&
\nonumber\\
&c\equiv\left(\frac{m_s}{H_*}\right)^2\sim 1.&
\label{Fs}
\eea
Here, $\Delta N$ is the number of the elapsed
e-foldings and $s_{\rm i}$ the initial value of
the inflaton field $s$. From the above, one can
obtain the inflation potential $N$ e-foldings
before the end of inflation as
\begin{equation}
V(N)\simeq V_{\rm m}\left(1-e^{-2F_sN}\right).
\label{VN}
\end{equation}

\par
Even though fast-roll, modular inflation keeps
the Hubble parameter $H$ rather rigid. Indeed,
it can be easily shown that
\begin{equation}
\epsilon=
\frac{1}{2}c^2\left(\frac{s}{M_{\rm P}}
\right)^2\ll 1,
\label{vareps}
\end{equation}
because \mbox{$c\sim 1$} and
\mbox{$s\ll M_{\rm P}$} during inflation with
$\epsilon$ being one of the so-called slow-roll
parameters defined as
\begin{equation}
\epsilon\equiv-\frac{\dot H}{H^2},
\label{SR}
\end{equation}
where the dot denotes derivative with
respect to the cosmic time.

\par
For modular inflation, the initial conditions
for the inflaton field are determined by the
quantum fluctuations which send the field off
the top of the potential hill. Hence, we expect
that the initial value for the inflaton is
\beq
s_{\rm i}\simeq\frac{H_{\rm m}}{2\pi},
\label{s0}
\eeq
where \mbox{$H_{\rm m}\simeq\sqrt{V_{\rm m}}/
\sqrt{3}M_{\rm P}$}.

\par
Using the above and considering that the final
value of $s$ is close to its vacuum expectation
value (VEV) \mbox{$s_{\rm VEV}\sim M_{\rm P}$},
we can estimate, through the use of
Eq.~(\ref{Fs}), the total number of e-foldings
as
\beq
N_{\rm tot}\simeq\frac{1}{F_s}
\ln\left(\frac{M_{\rm P}}{m_{3/2}}\right),
\label{Ntot}
\eeq
where we took into account that
\beq
H_{\rm m} \sim m_{3/2}.
\label{H0}
\eeq

\par
Inflation at such low energy scale as in
Eq.~(\ref{V0}) can provide the required
amplitude for the curvature perturbations only
through the use of a special kind of curvaton
field, whose perturbations are amplified during
inflation. In the following, we describe the
mechanism for such amplification.

\section{Amplifying the curvaton perturbations}
\label{sec:amplification}

\par
We discuss here the case of an axion-like
curvaton, i.e. a PNGB. Examples of such a
curvaton can be found in
Refs.~\cite{pngb,chun}. However, in contrast to
those works, we consider a PNGB curvaton whose
order parameter has \cite{amplif,lett} a
different (larger) expectation value in the
vacuum than during inflation and, in
particular, when the cosmological scales exit
the horizon. Thus, the potential for the real
canonically normalized curvaton field $\sigma$
is
\begin{eqnarray}
& &V(\sigma)=(v\tilde{m}_\sigma)^2\left[
1-\cos\left(\frac{\sigma}{v}\right)\right]
\nonumber \\
& &\Rightarrow\; V(|\sigma|\ll v)\simeq
\frac{1}{2}\tilde{m}_\sigma^2\sigma^2,
\label{Vs}
\end{eqnarray}
where \mbox{$v=v(t)$} is the order parameter
of the PNGB (determined by the values of the
radial fields in the model) with $t$ being the
cosmic time and \mbox{$\tilde{m}_\sigma=
\tilde{m}_\sigma(v)$} is the mass of the
curvaton at a given moment. In the true vacuum,
we have \mbox{$v=v_0$} and
\mbox{$\tilde{m}_\sigma=m_\sigma$} with $v_0$
and $m_\sigma$ being the order parameter and
the mass of the PNGB curvaton in the vacuum
respectively.

\par
Note that, in principle, the PNGB does not need
to have an exact sinusoidal potential. Instead,
one could substitute
\mbox{$[1-\cos(\sigma/v)]$} by $f(\sigma/v)$,
where $f$ is a periodic function of period
$2\pi$ with a global minimum at the origin and
$f(0)=0$. Then, generically, the second line in
Eq.~(\ref{Vs}) continues to be valid if $\sigma$
is close enough to the global minimum
\cite{footnote2}.

\subsection{The amplification factor}
\label{sec:factor}

\par
In this section, we will demonstrate that the
curvaton perturbations can be amplified by the
non-trivial evolution of its order parameter
$v$ if the curvaton is a PNGB. This mechanism
was first presented in Ref.~\cite{amplif}
(see also Ref.~\cite{lett}).

\par
We begin by using the fact that, on a
foliage of spacetime corresponding to spatially
flat hypersurfaces, the curvature perturbation
attributed to each of the universe components
(labelled by the index $i$) is given by
\cite{luw}
\begin{equation}
\zeta_i\equiv-H\frac{\delta\rho_i}
{\dot{\rho}_i},
\label{zetai}
\end{equation}
where $\rho_i$ and $\delta\rho_i$ are,
respectively, the energy density and its
perturbation of the component in question.

\par
The total curvature perturbation $\zeta(t)$,
which is also given by Eq.~(\ref{zetai}) with
$\rho_i$ and $\delta\rho_i$ replaced,
respectively, by the total energy density of
the universe $\rho=\sum_i\rho_i$ and its
perturbation $\delta\rho$, may be calculated
as follows. Using the fact that
\mbox{$\delta\rho=\sum_i\delta\rho_i$} and the
continuity equation
\mbox{$\dot{\rho}_i=-3H(\rho_i+p_i)$}, where
$p_i$ is the pressure of the $i$-th component
of the universe, it is easy to find that
\begin{equation}
\zeta=\sum_i\frac{\rho_i+p_i}{\rho+p}\;\zeta_i,
\label{zeta0}
\end{equation}
where $p=\sum_ip_i$ is the total pressure. Now,
since in the curvaton scenario, all
contributions to the curvature perturbation
other than the curvaton's are negligible, we
find that
\begin{equation}
\zeta=\zeta_\sigma
\left(\frac{1+w_\sigma}{1+w}\right)_{\rm dec}
\left.\frac{\rho_\sigma}{\rho}
\right|_{\,\rm dec},
\label{z}
\end{equation}
where \mbox{$\zeta\simeq 2\times 10^{-5}$} is
the curvature perturbation observed by the
cosmic microwave background explorer (COBE)
\cite{cobe}, $w_\sigma$ and $w$ are the
curvaton and the overall barotropic parameters
respectively (with
\mbox{$w_i\equiv p_i/\rho_i$}) and
$\zeta_\sigma$ is the partial curvature
perturbation of the curvaton. The right hand
side (RHS) of this equation is evaluated at the
time when the curvaton decays and this is
indicated by the subscript ``dec''. This decay
occurs after the end of inflation in which case
Eq.~(\ref{z}) gives
\begin{equation}
\zeta\sim\Omega_{\rm dec}\zeta_\sigma,
\label{zeta}
\end{equation}
where \mbox{$\Omega_{\rm dec}$} is the ratio of
the curvaton energy density to the total energy
density of the universe at the time of the
decay of the curvaton:
\begin{equation}
\Omega_{\rm dec}\equiv
\left.\frac{\rho_\sigma}{\rho}\right|_{\rm dec}
\leq 1.
\label{r}
\end{equation}
From the bound \cite{nongauss} on the possible
non-Gaussian component of the curvature
perturbation from the recent cosmic microwave
background radiation (CMBR) data obtained by
the Wilkinson microwave anisotropy probe (WMAP)
satellite, one finds \cite{luw} that, at $95\%$
confidence level (c.l.),
\beq
10^{-2}\lesssim\Omega_{\rm dec}\leq 1.
\label{wmap}
\eeq
The partial curvature perturbation of the
curvaton when the latter oscillates in a
quadratic potential [cf. Eq.~(\ref{Vs})] is
given \cite{CD} by
\begin{equation}
\zeta_\sigma=\frac{2}{3}
\left.\frac{\delta\sigma}{\sigma}
\right|_{\rm dec}\sim
\left.\frac{\delta\sigma}{\sigma}
\right|_{\rm osc},
\label{zs1}
\end{equation}
where ``osc'' denotes the time when the
curvaton oscillations begin.

\par
\begin{figure}[tb]
\centering
\includegraphics[width=\linewidth]{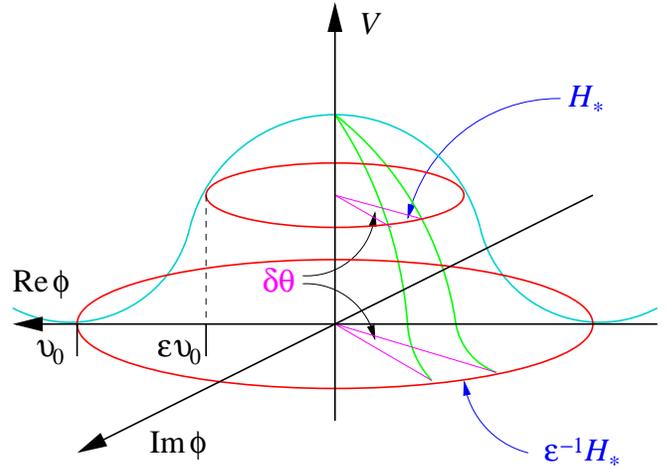}
\caption{Schematic representation of the
amplification of the PNGB curvaton
perturbation when the order parameter $v$
increases from the value it has when the
cosmological scales exit the horizon
\mbox{$v_*=\varepsilon v_0$} to its vacuum
value $v_0$. $V$ is the potential of a fiducial
complex field $\phi$ parametrizing the growth
of the order parameter such that $v/\sqrt{2}=
|\phi|$. The phase $\theta$ of $\phi$
corresponds to the PNGB degree of freedom
$\sigma$ (i.e. $\theta=\sigma/v$). The
perturbation in $\sigma$ at horizon crossing
has amplitude \mbox{$\delta\sigma_*\sim H_*$},
which corresponds to a phase perturbation of
magnitude \mbox{$\delta\theta=\delta\sigma_*/
v_*$}. As the order parameter grows,
$\delta\theta$ remains constant (the phase
perturbation is frozen on superhorizon scales)
but the amplitude of the curvaton perturbation
is increased up to
\mbox{$\delta\sigma\sim\varepsilon^{-1}H_*$}.}
\label{fig0}
\end{figure}

In this paper, we assume that the Hubble
parameter during inflation is comparable to
the (tachyonic) masses that the radial fields
which determine the value of the order
parameter of the PNGB acquire after inflation.
This means that the evolution of the curvaton's
order parameter $v$ ceases at (or soon after)
the end of inflation. Therefore, at the end of
inflation, \mbox{$v\rightarrow v_0$} and the
mass of the curvaton assumes its vacuum value
$m_\sigma$. Hence, in the following, {\em we
assume that the curvaton mass has already
assumed its vacuum value before the onset of
the curvaton oscillations}. Consequently, the
curvaton oscillations begin when
\begin{equation}
H_{\rm osc}\simeq\frac{1}{\sqrt 3}\; m_\sigma.
\label{Hosc}
\end{equation}
Before the oscillations begin, the phase
corresponding to the curvaton degree of freedom
is overdamped and remains frozen. More
precisely, this means that
\beq
\theta_{\rm osc}\simeq\theta_*,\quad
\delta\theta_{\rm osc}\simeq\delta\theta_*,
\label{osc*}
\eeq
where the subscript star denotes the values of
quantities at the time when the cosmological
scales exit the horizon during inflation,
\begin{equation}
\theta\equiv\frac{\sigma}{v}
\label{theta}
\end{equation}
with \mbox{$\theta\in(-\pi,\pi]$} and
\mbox{$\delta\theta$} is its perturbation.
Hence, for the curvaton partial perturbation,
we find
\begin{equation}
\left.\frac{\delta\sigma}{\sigma}
\right|_{\rm osc}=\left.\frac{\delta\theta}
{\theta}\,\right|_{\rm osc}\simeq
\left.\frac{\delta\theta}{\theta}\,\right|_*
=\left.\frac{\delta\sigma}{\sigma}\right|_*.
\label{fractional}
\end{equation}
Now, for the perturbation of the curvaton, we
have
\begin{equation}
\delta\sigma_*=\frac{H_*}{2\pi}.
\label{dsH}
\end{equation}
We assume that the order parameter of the PNGB
during inflation is smaller compared to its
value in the vacuum by a factor
\begin{equation}
\varepsilon\equiv\frac{v_*}{v_0}\ll 1.
\label{eps}
\end{equation}
Combining Eqs.~(\ref{osc*})-(\ref{eps}), we
find that
\begin{equation}
\delta\sigma_{\rm osc}\simeq
\frac{H_*}{2\pi\varepsilon},
\label{dsosc}
\end{equation}
which means that after the end of inflation,
when $v$ assumes its vacuum value, {\em
the curvaton perturbation is amplified by a
factor} $\varepsilon^{-1}$
(see Fig.~\ref{fig0}). From
Eqs.~(\ref{zeta}) and (\ref{zs1}), we have
\begin{equation}
\sigma_{\rm osc}\sim\frac{\Omega_{\rm dec}}
{\zeta}\,\delta\sigma_{\rm osc}.
\label{sigmaosc}
\end{equation}
Using Eq.~(\ref{dsosc}), we can recast the
above as
\begin{equation}
\sigma_{\rm osc}\sim
\frac{H_*\Omega_{\rm dec}}
{\pi\varepsilon\zeta}.
\label{sosc}
\end{equation}
We may obtain a lower bound on $\varepsilon$ as
follows:
\begin{equation}
\frac{\delta\sigma_*}{\sigma_*}\leq 1
\quad\Rightarrow\quad
\varepsilon\geq\varepsilon_{\rm min}
\equiv\frac{H_*}{2\pi v_0},
\label{epsbound}
\end{equation}
where we have used Eqs.~(\ref{fractional}) and
(\ref{dsosc}) and that
\mbox{$\sigma_{\rm osc}\lesssim v_0$}.

\subsection{The bounds on the inflationary
scale}
\label{sec:bounds}

\par
As is shown in Ref.~\cite{amplif}, in the case
when the curvaton oscillations begin after the
order parameter of the PNGB has attained its
vacuum value, we have
\bea
H_*&\sim&\Omega_{\rm dec}^{-\frac{2}{5}}
\left(\frac{H_*}{\min\{m_\sigma,
\Gamma_{\rm inf}\}}\right)^{\frac{1}{5}}
(\pi\varepsilon\zeta)^{\frac{4}{5}}
\nonumber \\
& &\times\left(\frac{\max\{H_{\rm dom},
\Gamma_\sigma\}}{H_{\rm BBN}}
\right)^{\frac{1}{5}}
(M_{\rm P}^3T_{\rm BBN}^2)^{\frac{1}{5}},
\label{H*}
\eea
or equivalently (using
\mbox{$V_*^{1/4}\simeq\sqrt{3^{1/2}H_*M_{\rm P}}$})
\bea
V_*^{\frac{1}{4}}&\sim&
\Omega_{\rm dec}^{-\frac{1}{5}}
\left(\frac{H_*}{\min\{m_\sigma,
\Gamma_{\rm inf}\}}\right)^{\frac{1}{10}}
(\pi\varepsilon\zeta)^{\frac{2}{5}}
\nonumber \\
& &\times\left(\frac{\max\{H_{\rm dom},
\Gamma_\sigma\}}{H_{\rm BBN}}
\right)^{\frac{1}{10}}
(M_{\rm P}^4T_{\rm BBN})^{\frac{1}{5}},
\label{V*}
\eea
where $\Gamma_{\rm inf}$ and $\Gamma_\sigma$
are the decay rates of the inflaton and the
curvaton fields respectively, $H_{\rm dom}$ is
the Hubble parameter at the time when the
curvaton energy density dominates the universe
(if the curvaton does not decay earlier), and
\mbox{$H_{\rm BBN}=(\pi/3)\sqrt{g_{\rm BBN}/10}
\;(T_{\rm BBN}^2/M_{\rm P})$} and
\mbox{$T_{\rm BBN}\approx 1~{\rm MeV}$} are,
respectively, the Hubble parameter and the
cosmic temperature at the time of big bang
nucleosynthesis (BBN) with
\mbox{$g_{\rm BBN}=10.75$} being the effective
number of relativistic degrees of freedom at
that time.

\par
Now, we require that the curvaton field decays
before BBN, i.e.
\mbox{$\Gamma_\sigma>H_{\rm BBN}$}. We also
have \mbox{$\Gamma_{\rm inf}\leq H_*$}. Hence,
Eqs.~(\ref{H*}) and (\ref{V*}) provide the
following bounds
\beq
H_*>\Omega_{\rm dec}^{-\frac{2}{5}}
(\pi\varepsilon\zeta)^{\frac{4}{5}}
(M_{\rm P}^3T_{\rm BBN}^2)^{\frac{1}{5}}
\!\sim\left(\frac{\varepsilon^2}
{\Omega_{\rm dec}}
\right)^{\frac{2}{5}}\times 10^7~{\rm GeV}
\label{H*bound}
\eeq
and
\beq
V_*^{\frac{1}{4}}>
\Omega_{\rm dec}^{-\frac{1}{5}}
(\pi\varepsilon\zeta)^{\frac{2}{5}}
(M_{\rm P}^4T_{\rm BBN})^{\frac{1}{5}}
\!\sim\left(\frac{\varepsilon^2}
{\Omega_{\rm dec}}\right)^{\frac{1}{5}}
\times 10^{12}~{\rm GeV}.
\label{V*bound}
\eeq

\par
Furthermore, we also note that, generically,
\begin{equation}
\Gamma_\sigma\geq\frac{m_\sigma^3}
{M_{\rm P}^2},
\label{Ggrav}
\end{equation}
where the equality corresponds to gravitational
decay. The above can be shown \cite{amplif} to
imply that
\begin{equation}
H_*\geq\Omega_{\rm dec}^{-1}
(\pi\varepsilon\zeta)^2M_{\rm P}
\left(\frac{m_\sigma}{H_*}\right)
\left(\max\left\{1,
\frac{m_\sigma}{\Gamma_{\rm inf}}
\right\}\right)^{\frac{1}{2}},
\label{Hbound1}
\end{equation}
which results in the bounds
\bea
H_*&\geq&\Omega_{\rm dec}^{-1}
(\pi\varepsilon\zeta)^2M_{\rm P}
\left(\frac{m_\sigma}{H_*}\right)
\nonumber \\
& &\sim\left(\frac{\varepsilon^2}
{\Omega_{\rm dec}}\right)
\left(\frac{m_\sigma}{H_*}\right)
\times 10^{10}~{\rm GeV}
\label{Hbound}
\eea
and
\bea
V_*^{\frac{1}{4}}&\geq&
\Omega_{\rm dec}^{-\frac{1}{2}}
(\pi\varepsilon\zeta)\;M_{\rm P}
\left(\frac{m_\sigma}{H_*}\right)^{\frac{1}{2}}
\nonumber \\
& &\sim
\left(\frac{\varepsilon^2}{\Omega_{\rm dec}}
\right)^{\frac{1}{2}}
\left(\frac{m_\sigma}{H_*}\right)^{\frac{1}{2}}
\times 10^{14}~{\rm GeV}.
\label{Vbound}
\eea
These bounds may be relaxed if $\varepsilon$ is
small enough. For a PNGB curvaton in particular,
we may have \mbox{$m_\sigma\ll H_*$}, which
also relaxes the bounds in Eqs.~(\ref{Hbound})
and (\ref{Vbound}). However, in our case,
\mbox{$m_\sigma\sim H_*$} (see below).
Comparing the bounds in Eqs.~(\ref{H*bound})
and (\ref{V*bound}) with those in
Eqs.~(\ref{Hbound}) and (\ref{Vbound})
respectively, we find that the first set of
bounds is more stringent if
\bea
\varepsilon&<&
\frac{1}{\pi\zeta
\Omega^{\frac{1}{2}}_{\rm dec}}\left(
\frac{T_{\rm BBN}}{M_{\rm P}}
\right)^{\frac{1}{3}}
\left(\frac{H_*}{m_\sigma}\right)^{\frac{5}{6}}
\nonumber \\
& &\sim
10^{-3}\,\Omega_{\rm dec}^{-\frac{1}{2}}
\left(\frac{H_*}{m_\sigma}
\right)^{\frac{5}{6}}.
\label{ebound}
\eea
Thus, for \mbox{$\varepsilon\ll 1$}, the second
set of bounds is typically less stringent than
the first one.

\par
From Eqs.~(\ref{epsbound}) and
(\ref{V*bound}) and after a little
algebra, it is easy to get
\bea
& &V_*^{\frac{1}{4}}\geq\left(\frac{M_{\rm P}}
{v_0}\right)^2 10^{-13}~{\rm GeV}
\nonumber \\
& &\Rightarrow\; H_* \geq
\left(\frac{M_{\rm P}}{v_0}\right)^4
10^{-44}~{\rm GeV},
\label{Vvbound}
\eea
which means that, in principle,
{\em the larger $v_0$ is the
smaller $V_*^{1/4}$ can become}.

\subsection{Scale invariance requirement}
\label{sec:ns}

\par
The evolution of the order parameter $v(t)$
during inflation is subject to an important
constraint which has to do with preserving the
scale invariance of the spectrum of the
curvature perturbations.

\par
The amplitude of the curvature perturbation is
determined by the magnitude of the
perturbation in the curvaton field, which, in
this scenario, apart from the scale of $H_*$ is
also determined by the amplification factor
$\varepsilon^{-1}$. The latter is determined by
the value of the order parameter $v_*$ when the
curvaton quantum fluctuations exit the horizon
during inflation. A strong variation of $v(t)$
at that time results in a strong dependence of
$\varepsilon(k)$ on the comoving momentum scale
$k$, which would reflect itself on the
perturbation spectrum threatening significant
departure from scale invariance.

\par
In Ref.~\cite{amplif}, it was shown that, in
order for this to be avoided, the rate of
change of the order parameter must be
constrained as
\begin{equation}
\left|\left(\frac{\dot v}{v}\right)_*
\right|\ll H_*.
\label{nsbound}
\end{equation}
From the above, it is evident that, in order
not to violate the observational constraints
regarding the scale invariance of the curvature
perturbation spectrum, {\em the order parameter
must either remain constant or, at most, have a
very slow variation when the cosmological
scales exit the horizon} \cite{footnote3}.
However, this cannot remain so indefinitely
because we need \mbox{$v_0\gg v_*$} to have
substantial amplification of the perturbation
(i.e. \mbox{$\varepsilon\ll 1$}). Consequently,
$v$ has to increase dramatically at some point
{\em after} the exit of the cosmological
scales from the inflationary horizon. This
requirement is crucial for model building.

\par
In this paper, we will show that the evolution
of $v$ begins at a phase transition during
inflation, as presented in Ref.~\cite{lett}.
Initially, the growth of $v$ is very slow, but
later, near the end of inflation, $v$ grows
substantially until it reaches its vacuum value
$v_0$.

\section{Can we construct a PQ model with a
PNGB curvaton?}
\label{sec:quest}

\par
We will now address the question whether we can
construct a PQ model \cite{pq}
which, in addition to the standard axion,
contains another axion-like field with the
right properties to play the role of a PNGB
curvaton as discussed in the previous section.
It is well known that, in the SUGRA
extension of the minimal supersymmetric
standard model (MSSM), there exist certain D-
and F-flat directions in field space which can
generate intermediate scales
\begin{equation}
M_{\rm I}\sim (m_{3/2}
M_{\rm P}^n)^{\frac{1}{n+1}},
\label{eq:inter}
\end{equation}
where $n$ is a positive integer. It seems
natural to try to identify $M_{\rm I}$ with the
symmetry breaking scale $f_{\rm a}$ of the PQ
symmetry ${\rm U}(1)_{\rm PQ}$, such that a
$\mu$ term is generated with $\mu\sim
f^{n+1}_{\rm a}/M_{\rm P}^n\sim m_{3/2}$
\cite{kn}. This would simultaneously resolve
the strong $CP$ and $\mu$ problems of MSSM.

\par
The resolution of the $\mu$ problem forces us
to consider non-renormalizable superpotential
terms such as
\beq
\lambda P^{n+1}h_1h_2/M_{\rm P}^n,
\label{eq:muterm}
\eeq
where $\lambda$ is a dimensionless parameter,
$P$ is a SM singlet superfield
and $h_1$, $h_2$ are the electroweak Higgs
doublets. The fact that the PQ symmetry carries
QCD anomalies implies that the combination
$h_1h_2$ must have a non-zero PQ charge as one
can easily deduce from the Yukawa couplings of
the quarks. The field $P$ must then necessarily
carry a non-zero PQ charge. So, if it acquires
a VEV of order $M_{\rm I}$, the PQ symmetry
breaks spontaneously and a $\mu$ term of the
right magnitude is generated via the
superpotential term in Eq.~(\ref{eq:muterm}).

\par
However, the field $P$ has no self-couplings
due to its non-zero PQ charge. Moreover, its
couplings to the MSSM fields involve at least
two of them since there are no SM singlets in
MSSM. As a consequence, before the electroweak
symmetry breaking, $P$ has a flat potential. To
lift the flatness of its potential and generate
an intermediate VEV for $P$ of the order of
$M_{\rm I}$ in Eq.~(\ref{eq:inter}), we must
introduce \cite{choi,rsym,thermal} a second SM
singlet superfield $Q$ with non-zero PQ charge
having a coupling of the type
\beq
\xi P^{n+3-k}Q^k/M_{\rm P}^n,
\label{eq:PQcoupling}
\eeq
where $\xi$ is a dimensionless coupling
constant and $k$ is a positive integer smaller
than $n+3$. The superpotential term in
Eq.~(\ref{eq:PQcoupling}) determines the PQ
charge of $Q$.

\par
After soft SUSY breaking, we obtain the
following scalar potential for the spontaneous
breaking of the PQ symmetry:
\bea
V&=&|\xi|^2\left[(n+3-k)^2|Q|^2+k^2|P|^2\right]
\left|\frac{P^{n+2-k}Q^{k-1}}{M_{\rm P}^n}
\right|^2
\nonumber \\
& &+m_P^2|P|^2+m_Q^2|Q|^2
\nonumber \\
& &+\left[A\xi\frac{P^{n+3-k}Q^k}{M_{\rm P}^n}+
{\rm h.c.}\right],
\label{VPQ}
\eea
where $m_P^2$, $m_Q^2\sim m_{3/2}^2$ and can
have either sign, while $A$ is a complex
parameter with magnitude of order $m_{3/2}$.
For large enough $|A||\xi|$, this potential
possesses non-trivial (local) minima at
\beq
|P|,|Q|\sim (m_{3/2}
M_{\rm P}^n)^{\frac{1}{n+1}}
\label{PQscale}
\eeq
even if $m_P^2$, $m_Q^2$ are positive since the
last term in the RHS of Eq.~(\ref{VPQ}) can be
adequately negative. We see that here the PQ
scale $f_{\rm a}$ is generated dynamically and
is not inserted by hand as in the PQ schemes
with renormalizable interactions.

\par
In order to implement our scenario, we need a
valley of local minima of the potential (with
respect to the direction perpendicular to the
valley) which has negative inclination. Along
this valley, the fields $|P|$ and $|Q|$ must
take values which are much smaller than their
vacuum values. If the system happens to slowly
roll down this valley during the relevant part
of inflation, the order parameter $v$ of the
PNGB remains, during inflation, much smaller
than its vacuum value $v_0$ and our
amplification mechanism for the curvaton
perturbations may work. This can be achieved
only if one of the masses-squared $m_P^2$,
$m_Q^2$ is negative. Let us assume, for
definiteness, that $m_P^2<0$ and $m_Q^2>0$.
Note, however, that the following discussion
applies equally well to the opposite case too,
where $m_P^2>0$ and $m_Q^2<0$. In the case
under consideration, the scalar potential is
\cite{thermal} unbounded below on the $P$ axis
(i.e. for $Q=0$) unless $k=1$ since, for $k>1$,
all the terms in the RHS of Eq.~(\ref{VPQ})
vanish on the $P$ axis except the negative mass
term of $P$. So, we restrict ourselves to the
case $k=1$.

\par
During inflation, $m_P^2$, $m_Q^2$ acquire
\cite{crisis,cllsw,randall} SUGRA corrections
of order $H^2$ which are assumed to be positive.
Also, $A$ receives SUGRA corrections of order
$H$. As already mentioned, the Hubble parameter
during inflation is, in our case, of order
$m_{3/2}$. So, in the initial stages of
inflation, the effective mass-squared of $P$
after SUGRA corrections, which we
will call $\bar{m}_P^2$, can be positive. In
this case, the origin in field space becomes a
local minimum and the system may be initially
trapped there. As $H$ gradually decreases
during inflation, $\bar{m}_P^2$ becomes
negative and the system starts slowly rolling
down in the $P$ direction. For non-zero $P$,
the last term in the RHS of Eq.~(\ref{VPQ})
yields a linear term in $Q$ (recall that $k=1$)
and, thus, $Q$ is also shifted from zero. More
precisely, we get
\beq
|Q|\sim \frac{|P|^{n+1}}{m_{3/2}M_{\rm P}^n}|P|
\ll |P|.
\label{|P||Q|}
\eeq

\par
The last term in the RHS of Eq.~(\ref{VPQ})
then yields
\beq
-2|\bar{A}||\xi|\frac{|P|^{n+2}|Q|}
{M_{\rm P}^n}
\cos\left[\frac{(n+2)\phi_P}{\sqrt{2}|P|}+
\frac{\phi_Q}{\sqrt{2}|Q|}\right],
\label{APQterm}
\eeq
where $\phi_P$ and $\phi_Q$ are canonically
normalized real fields corresponding to the
phases of $P$ and $Q$ respectively and
$\bar{A}$ is the effective $A$ after SUGRA
corrections. Note that,
in deriving Eq.~(\ref{APQterm}), the fields
$P$ and $Q$ were appropriately rephased so that
the product $\bar{A}\xi$ is negative. This is a
convenient choice since, in this case, the
expression in the above equation is minimized
when the argument of the cosine vanishes. The
orthogonal axion direction, which we would like
to use as a PNGB curvaton, corresponds to the
canonically normalized real field
\beq
\frac{(n+2)|Q|\phi_P+|P|\phi_Q}
{\sqrt{(n+2)^2|Q|^2+|P|^2}}.
\label{ortho}
\eeq
Its mass-squared can be evaluated from the term
in Eq.~(\ref{APQterm}), which, in view of
Eq.~(\ref{|P||Q|}), yields
\beq
\frac{|\bar{A}||\xi||P|^{n+2}}{M_{\rm P}^n|Q|}
\sim m_{3/2}^2.
\label{orthomass}
\eeq
So, the mass of the orthogonal axion during
inflation is of order $m_{3/2}$, which is
comparable to the inflationary Hubble
parameter. Consequently, this field does not
qualify as a curvaton, because it cannot
obtain a superhorizon spectrum of
perturbations. In conclusion, we
have seen that our scenario cannot be realized
within extensions of the MSSM with a PQ
symmetry which contain only two SM singlet
superfields.

\par
The addition of a third SM singlet superfield
$S$, however, can drastically change the
situation allowing the implementation of our
mechanism. We could keep the masses-squared of
$P$ and $Q$ positive and include a
superpotential term of the type in
Eq.~(\ref{eq:PQcoupling}) with any value of
the integer $k$ between unity and $n+2$. In
this case, as already mentioned, the potential
for $P$ and $Q$ (with $S=0$) can possess
non-trivial minima at $|P|$ and $|Q|$ given
by Eq.~(\ref{PQscale}). Now, we introduce an
extra superpotential term of the type
\beq
\xi_q P^{n+3-p-q}Q^pS^q/M_{\rm P}^n,
\label{eq:PQScoupling}
\eeq
where $p$, $q$ are non-negative integers with
$p+q\leq n+3$ and $q\geq 3$. This term
determines the PQ charge of $S$. Of course, we
should keep in mind that all possible terms
involving $P$, $Q$ and $S$ that satisfy the
global symmetries (including R symmetries) of
the terms in Eqs.~(\ref{eq:PQcoupling}) and
(\ref{eq:PQScoupling}) should be present in the
superpotential. We assume that the term in
Eq.~(\ref{eq:PQScoupling}) is the term of this
type with the smallest power of $S$. We take
the mass-squared of $S$ negative. Then, for
small values of $S$, we obtain a valley of
minima with negative inclination at almost
constant values of $|P|$ and $|Q|$ given by
Eq.~(\ref{PQscale}).

\par
If the system slowly rolls down this valley
during inflation, the $A$-term corresponding to
the coupling in Eq.~(\ref{eq:PQcoupling})
generates a mass term for the canonically
normalized field
\beq
\frac{(n+3-k)|Q|\phi_P+k|P|\phi_Q}
{\sqrt{(n+3-k)^2|Q|^2+k^2|P|^2}}
\label{orthon}
\eeq
with mass-squared of order $m_{3/2}^2$. This is
easily shown by repeating the argument which led
to Eqs.~(\ref{ortho}) and (\ref{orthomass}). The
$A$-term corresponding to the coupling in
Eq.~(\ref{eq:PQScoupling}) yields
\bea
&-&2|\bar{A}||\xi_q|
\frac{|P|^{n+3-p-q}|Q|^p|S|^q}{M_{\rm P}^n}
\nonumber\\
&\times&\cos\left[\frac{(n+3-p-q)\phi_P}
{\sqrt{2}|P|}+\frac{p\phi_Q}
{\sqrt{2}|Q|}+\frac{q\phi_S}{\sqrt{2}|S|}
\right]\!,
\label{APQSterm}
\eea
where $\phi_S$ is a canonically normalized real
scalar field corresponding to the phase of $S$
and $\bar{A}\xi_q$ was taken negative by
rephasing $S$. This generates a ``mass term''
for the canonically normalized field
\beq
\frac{(n+3-p-q)|Q||S|\phi_P+p|P||S|\phi_Q+
q|P||Q|\phi_S}{\sqrt{(n+3-p-q)^2|Q|^2|S|^2+
p^2|P|^2|S|^2+q^2|P|^2|Q|^2}}
\label{orthoS}
\eeq
with ``mass-squared''
\beq
\frac{q^2|\bar{A}||\xi_q||P|^{n+3-p-q}|Q|^p
|S|^{q-2}}{M_{\rm P}^n}\sim m_{3/2}^2\left(
\frac{|S|}{|P|}\right)^{q-2}
\label{orthomassS}
\eeq
for $|S|\ll |P|\sim |Q|$ (we use quotation
marks to indicate that the ``mass term'' and
``mass-squared'' do not necessarily correspond
to a mass eigenstate). We see that, for
$q\geq 3$, this ``mass-squared'' is suppressed
relative to $m_{3/2}^2$. Superpotential terms
with higher powers of $S$ obviously give even
more suppressed ``masses-squared''. So, the
linear combination of $\phi_P$, $\phi_Q$ and
$\phi_S$ which is orthogonal to the combination
in Eq.~(\ref{orthon}) and the axion direction
has mass-squared much smaller than $m_{3/2}^2$
during inflation and can be used as PNGB
curvaton. It is though important to make sure
that at least one of the three terms
$S^{n+3}/M_{\rm P}^n$, $S^{n+2}P/M_{\rm P}^n$
and $S^{n+2}Q/M_{\rm P}^n$ is allowed in the
superpotential, since otherwise the potential
will be unbounded below on the $S$ axis. In the
next section, we will present a concrete class
of models of this category.

\section{PQ models with an axion-like curvaton}
\label{sec:model}

\par
We consider a class of simple extensions of
MSSM which are based on the SM gauge group
$G_{\rm SM}$, but also possess two continuous
global ${\rm U}(1)$ symmetries, namely a PQ
symmetry ${\rm U}(1)_{\rm PQ}$ and a R symmetry
${\rm U}(1)_{\rm R}$, and a discrete $Z_2^P$
symmetry. In addition to the usual MSSM
left-handed superfields $h_1$, $h_2$ (Higgs
${\rm SU}(2)_{\rm L}$ doublets), $l_i$
(${\rm SU}(2)_{\rm L}$ doublet leptons),
$e^c_i$ (${\rm SU}(2)_{\rm L}$ singlet
charged leptons), $q_i$ (${\rm SU}(2)_{\rm L}$
doublet quarks), and $u^c_i$, $d^c_i$
(${\rm SU}(2)_{\rm L}$ singlet anti-quarks)
with $i=1,2,3$ being the family index, the
models also contain the SM singlet left-handed
superfields $P$, $Q$ and $S$. The charges of
the superfields under ${\rm U}(1)_{\rm PQ}$ and
${\rm U}(1)_{\rm R}$ are
\begin{eqnarray}
{\rm PQ}:~P(-2),~Q(2),~S(0),~h_1,h_2(n+1),
\nonumber
\\
{\rm R}:~P(\frac{n+3}{2}),~Q(\frac{n-1}{2}),
~S(\frac{n+1}{2}),~h_1,h_2(0)
\label{charges}
\end{eqnarray}
with the ``matter'' (quark and lepton)
superfields having ${\rm PQ}=-(n+1)/2$,
${\rm R}=(n+1)(n+3)/4$. The integer $n$ is
taken to be of the form
\beq
n=4l+1,
\label{n4l1}
\eeq
where $l=0,1,2,...$ is any non-negative integer
providing a numbering of the models in this
class (see Sec.~\ref{sec:shift}), and the
charges are normalized so that they take their
absolutely smallest possible integer values.
Finally, under the $Z_2^P$ symmetry, $P$
changes sign.

\par
The most general superpotential compatible with
these symmetries is
\bea
W &=&
y_{eij}(l_ih_1)e^c_j+y_{uij}(q_ih_2)u^c_j+
y_{dij}(q_ih_1)d^c_j
\nonumber \\
& &+\lambda P^{n+1}(h_1h_2)/M_{\rm P}^n
\nonumber \\
& &+\sum_{k=0}^{(n+3)/4}\lambda_k
S^{n+3-4k}(PQ)^{2k}/M_{\rm P}^n,
\label{W}
\eea
where $y_{eij}$, $y_{uij}$, $y_{dij}$ are the
usual Yukawa coupling constants, $\lambda$,
$\lambda_k$ are complex dimensionless
parameters, $(X Y)$ indicates
the ${\rm SU}(2)_{\rm L}$ invariant product
$\eps_{ab}X_aY_b$ with $\eps$ denoting the
$2\times 2$ antisymmetric matrix with
$\eps_{12}=1$, and summation over the family
indices is implied. The R charge of $W$ is
(n+1)(n+3)/2. Baryon number is
automatically conserved to all orders in
perturbation theory as a consequence of the
R symmetry. The reason is
\cite{nonthtripletdec} that the R charge of
any combination of three color triplet or
anti-triplet superfields exceeds the R charge
of $W$ and there are no superfields with
negative R charge to compensate. Note that the
$Z_2$ subgroup of ${\rm U}(1)_{\rm PQ}$
coincides with the discrete matter parity
symmetry (denoted by $Z_2^{\rm mp}$), which
changes the sign of all matter superfields. It
is obvious that the superpotential in
Eq.~(\ref{W}) also conserves lepton number,
which is a consequence of both the R and the PQ
symmetry.

\par
Note that the superpotential in Eq.~(\ref{W})
is of the same type as the superpotential which
has been considered in Ref.~\cite{chun}. The
main difference is that here the discrete
$Z_{n+3}$ symmetry of Ref.~\cite{chun} is
replaced by a much more powerful continuous
${\rm U}(1)$ R symmetry and, also, that an
extra discrete $Z_2^P$ symmetry is added.
Moreover, in contrast to Ref.~\cite{chun}, we
include here all the superpotential terms which
are compatible with the symmetries of the
model. In this sense, the superpotential in
Eq.~(\ref{W}) is a completely natural
superpotential. It should be emphasized that
continuous global symmetries such as the
${\rm U}(1)_{\rm R}$ or the
${\rm U}(1)_{\rm PQ}$ symmetry used here,
rather than being imposed, can arise in a
natural manner from an underlying superstring
theory. Indeed, as it was pointed out in
Ref.~\cite{lps}, discrete symmetries (including
R symmetries) that typically arise after
compactification could effectively behave as if
they are continuous.

\par
To see that the above superpotential has the
most general form allowed by the symmetries,
observe that, due to matter parity, any term in
$W$ must contain an even number of matter
superfields. Actually, we may have either no or
two matter fields since higher combinations
carry R charge larger than that of $W$.
The possible combinations of two matter fields
are of the type $ll$, $e^ce^c$, $le^c$, $qu^c$
and $qd^c$ (from color conservation) and
have the R charge of $W$. So, we can multiply
them only by superfields of zero R charge,
i.e. $h_1$, $h_2$, and $Q$ in the case $n=1$.
However, multiplying $ll$, $e^ce^c$ just by
$Q$'s, we cannot compensate their non-zero weak
hypercharge. We need to multiply them at least
once by $h_1$ or $h_2$. Indeed, the weak
hypercharge of $ll$ ($e^ce^c$) is compensated
if we take the combination $llh_2h_2$
($e^ce^ch_1h_1h_1h_1$) whose PQ charge is n+1
(3(n+1)), which could only be cancelled by
including $P$'s, the only superfields with
negative PQ charge. This is though not allowed
by R symmetry. So, the only matter field
bilinears which are allowed are $qu^c$, $qd^c$
and $le^c$, which conserve lepton number.
Actually, to cancel all the SM quantum numbers
and the PQ charge, we must take the
combinations $(qh_2)u^c$, $(qh_1)d^c$ and
$(lh_1)e^c$. No further superfields can be
included in these combinations since, by
${\rm U}(1)_{\rm PQ}$, we could only include
the combination $PQ$, which has though positive
R charge. In conclusion, we see that the only
superpotential terms involving matter
superfields which can be present are the usual
Yukawa couplings.

\par
We still have to consider superpotential terms
with no matter superfields. If such terms
involve $h_1$, $h_2$, these superfield must
enter through the combination $(h_1h_2)$ which
is neutral under both ${\rm SU}(2)_{\rm L}$ and
${\rm U}(1)_Y$. To cancel the PQ charge, we
must then take $P^{n+1}(h_1h_2)$. To retain the
${\rm U}(1)_{\rm R}$ symmetry, we could
multiply this only by the combination $PQ$,
which has ${\rm PQ}=0$.
However, the R charge of $P^{n+1}(h_1h_2)$ is
already equal to the R charge of $W$ and $PQ$
has positive R charge. So, $P^{n+1}(h_1h_2)$ is
the only allowed term which involves Higgs but
no matter superfields. We are left to consider
combinations which involve only SM singlet
superfields. The PQ symmetry allows only the
combinations $S^p(PQ)^q$, where $p$ and $q$ are
non-negative integers. The R charge of $PQ$
($S$) is $n+1$ ($(n+1)/2$), which implies that
the term $S^{n+3}$ can be present in the
superpotential. Note that the combination $PQ$
has the same R charge as $S^2$. Consequently,
the terms $S^{n+1}(PQ)$, $S^{n-1}(PQ)^2$,...,
$(PQ)^{(n+3)/2}$ are also allowed by
${\rm U}(1)_{\rm R}$. The $Z_2^P$ symmetry,
however, forbids all the odd powers of $PQ$ and
we arrive at the terms in the sum in the RHS of
Eq.~(\ref{W}).

\par
The PQ symmetry is anomalous (as it should). In
particular, one can show that the QCD
instantons break it explicitly to its
$Z_\mathcal{N}$ subgroup with
$\mathcal{N}=6(n+1)$. This subgroup contains
the $Z_2^{\rm mp}$ (matter parity), which thus
always remains unbroken by instanton effects.
On the contrary, the R symmetry is
non-anomalous since the fermionic components of
all the color triplet or anti-triplet
superfields (actually, all the matter fermions)
have zero R charge. The soft
SUSY-breaking terms (especially, the
$A$-type terms), however, break
${\rm U}(1)_{\rm R}$ to its $Z_\mathcal{M}$
subgroup, where $\mathcal{M}=(n+1)(n+3)/2$.

\section{The scalar potential}
\label{sec:potential}

\par
The part of the superpotential in Eq.~(\ref{W})
which is relevant for the PQ breaking is the
sum in the RHS of this equation. This sum
contains at least two terms. For $n=1$, in
particular, it consists of just two terms. The
resulting scalar potential after including soft
SUSY-breaking effects is
\beq
V=|F_P|^2+|F_Q|^2+|F_S|^2+ V_{\rm soft},
\label{V}
\eeq
where
\beq
F_P=\sum_{k=1}^{(n+3)/4}2k\lambda_k
\frac{S^{n+3-4k}(PQ)^{2k-1}Q}{M_{\rm P}^n}
\equiv FQ,
\label{FP}
\eeq
\beq
F_Q=\sum_{k=1}^{(n+3)/4}2k\lambda_k
\frac{S^{n+3-4k}(PQ)^{2k-1}P}{M_{\rm P}^n}
\equiv FP
\label{FQ}
\eeq
and
\beq
F_S=\sum_{k=0}^{(n-1)/4}(n+3-4k)\lambda_k
\frac{S^{n+2-4k}(PQ)^{2k}}{M_{\rm P}^n}
\label{FS}
\eeq
are the F-terms, and
\bea
V_{\rm soft}&=&m_P^2|P|^2+m_Q^2|Q|^2+
m_S^2|S|^2
\nonumber \\
& &+\left[A\sum_{k=0}^{(n+3)/4}\lambda_k
\frac{S^{n+3-4k}(PQ)^{2k}}{M_{\rm P}^n}+
{\rm h.c.}\right]
\label{Vsoft}
\eea
the soft SUSY-breaking terms. Here, the soft
SUSY-breaking masses-squared $m_P^2$, $m_Q^2$
and $m_S^2$ are of the order of the $m_{3/2}^2$
and can have either sign. Also, for simplicity,
we assumed universal soft SUSY-breaking
$A$-terms with the magnitude of the complex
parameter $A$ being of the order of $m_{3/2}$.
Note that the sums in the RHS of
Eqs.~(\ref{FP})-(\ref{FS}) contain at least one
term. Actually, these sums consists of just one
term for $n=1$. On the other hand, the sum in
Eq.~(\ref{Vsoft}) contains at least two terms
with the minimum number of terms corresponding
to $n=1$.

\par
As $|P|$, $|Q|$, $|S|\rightarrow\infty$, the
potential is generally dominated by the
positive F-terms and thus $V$ is bounded below
independently of the sign of the soft mass
terms. However, on the $P$ or $Q$ axis,
i.e. for $Q=S=0$ or $P=S=0$ respectively, the
F-terms vanish identically together with the
$A$-terms and the potential is given by just
the mass term of $P$ or $Q$ respectively. So,
to have $V$ bounded below on these two axes
too, we must restrict the masses-squared of $P$
and $Q$ to be positive. For simplicity, we will
take these two soft masses-squared to be equal,
i.e. we will put
\beq
m_P^2=m_Q^2\equiv m^2.
\label{mPmQm}
\eeq
The potential on the $S$ axis (i.e. for
$P=Q=0$) is
\beq
V=(n+3)^2|\lambda_0|^2\frac{|S|^{2(n+2)}}
{M_{\rm P}^{2n}}+m_S^2|S|^2+\left(A\lambda_0
\frac{S^{n+3}}{M_{\rm P}^n}+{\rm h.c.}\right)
\label{Sdir}
\eeq
with the first term in the RHS of this equation
originating from the F-term $F_S$. We see that,
for $|S|\rightarrow\infty$, the potential is
dominated by the positive F-term and thus $V$
on the $S$ axis is bounded below no matter
what the sign of the mass-squared of $S$ is.
So, this sign can be chosen at will. For
reasons which will become clear later, however,
we take
\beq
m_S^2<0.
\label{massSneg}
\eeq
Therefore, the origin in field space
($P=Q=S=0$) is a saddle point of the potential
with positive curvature in the $P$ and $Q$
directions and negative in the $S$ direction.

\par
Using Eqs.~(\ref{V})-(\ref{Vsoft}), one can
readily show that, for $m_P^2$, $m_Q^2>0$, the
potential $V$ in Eq.~(\ref{V}) has a valley of
local minima with respect to $|P|$ and $|Q|$
which lies on the $S$ axis (i.e. at
$P=Q=0$). The potential right on the bottom
line of this valley can be found from
Eq.~(\ref{Sdir}) by choosing the phase
$\theta_S$ of $S$ so that the sum of the terms
in the parentheses in the RHS of
Eq.~(\ref{Sdir}) is minimized. In this case,
this equation takes the form
\beq
V=(n+3)^2|\lambda_0|^2\frac{|S|^{2(n+2)}}
{M_{\rm P}^{2n}}+m_S^2|S|^2-2|A||\lambda_0|
\frac{|S|^{n+3}}{M_{\rm P}^n},
\label{Sdir||}
\eeq
which has a minimum at
\beq
\frac{|S|^{n+1}}{M_{\rm P}^n}=
\frac{|A|+\sqrt{|A|^2-4(n+2)m_S^2}}
{2(n+2)(n+3)|\lambda_0|},
\label{trmin}
\eeq
where
$|S|\sim (m_{3/2}M_{\rm P}^n)^{\frac{1}{n+1}}$.
So, as $|S|$ increases from zero, the depth of
the valley increases (i.e. the valley has
initially negative inclination) until $|S|$
reaches the value in Eq.~(\ref{trmin}), where
the maximal depth is achieved. As $|S|$
increases further, the bottom line of the
valley rises and, for $|S|\rightarrow\infty$,
tends to infinity. We will call this valley,
which starts from the trivial saddle point at
the origin, the trivial valley.

\subsection{The non-trivial minima of the
potential}
\label{sec:minima}

\par
The extrema of the full potential $V$ in
Eq.~(\ref{V}) are given by the equations
\beq
\frac{\partial V}{\partial P}=F_P^*
\frac{\partial F_P}{\partial P}+F_Q^*
\frac{\partial F_Q}{\partial P}+F_S^*
\frac{\partial F_S}{\partial P}+m^2P^*+
AF_P=0,
\label{fullPder}
\eeq
\beq
\frac{\partial V}{\partial Q}=F_P^*
\frac{\partial F_P}{\partial Q}+F_Q^*
\frac{\partial F_Q}{\partial Q}+F_S^*
\frac{\partial F_S}{\partial Q}+m^2Q^*+
AF_Q=0
\label{fullQder}
\eeq
and
\beq
\frac{\partial V}{\partial S}=F_P^*
\frac{\partial F_P}{\partial S}+F_Q^*
\frac{\partial F_Q}{\partial S}+F_S^*
\frac{\partial F_S}{\partial S}+m^2_SS^*+
AF_S=0.
\label{fullSder}
\eeq
Multiplying Eqs.~(\ref{fullPder}) and
(\ref{fullQder}) by $P$ and $Q$ respectively,
subtracting and using
Eqs.~(\ref{FP})-(\ref{FS}), we obtain
\beq
(|F|^2+m^2)(|P|^2-|Q|^2)=0,
\label{eq|P||Q|}
\eeq
which implies that $|P|=|Q|$. So, the complex
fields $P$ and $Q$ have exactly the same
magnitude in any extremum of the full potential
$V$. This is actually true also in the extrema
of $V$ with respect to $P$ and $Q$ only for
fixed $S$ since we have not used
Eq.~(\ref{fullSder}). In the trivial minimum in
Eq.~(\ref{trmin}) which lies on the trivial
valley, $|P|=|Q|=0$. However, the potential $V$
possesses non-trivial minima too, where
\beq
|P|=|Q|, |S|\sim
(m_{3/2}M_{\rm P}^n)^{\frac{1}{n+1}}.
\label{nontrmin}
\eeq

\par
It is important to note that the continuous
global symmetry of the model is not exactly
${\rm U}(1)_{\rm PQ}\times {\rm U}(1)_{\rm R}$.
The reason is that there are elements of
${\rm U}(1)_{\rm R}$ which are
indistinguishable from elements of
${\rm U}(1)_{\rm PQ}$ since they have the same
action on all the superfields of the model.
Actually, the only elements of
${\rm U}(1)_{\rm R}$ which can, in principle,
be identified with elements of
${\rm U}(1)_{\rm PQ}$ are the ones belonging to
its $Z_\mathcal{M}$ subgroup with
$\mathcal{M}=(n+1)(n+3)/2$ generated by the
element
\beq
e^{i\frac{2\pi}{\mathcal{M}}}\in
{\rm U}(1)_{\rm R}.
\label{ZM}
\eeq
This is so because this $Z_\mathcal{M}$ is the
maximal ordinary (i.e. non-R) symmetry group
contained in ${\rm U}(1)_{\rm R}$. One can show
that the element
\beq
e^{-i\frac{8\pi}{n+1}}\in {\rm U}(1)_{\rm R}
\label{Z(n+1)/2R}
\eeq
has the same action on all the superfields as
the element
\beq
e^{i\frac{4\pi}{n+1}}\in {\rm U}(1)_{\rm PQ}
\label{Z(n+1)/2PQ}
\eeq
and, thus, the $Z_{(n+1)/2}$ subgroup of
$Z_\mathcal{M}$ which is generated by it is
identical with the $Z_{(n+1)/2}$ subgroup
of ${\rm U}(1)_{\rm PQ}$ generated by the
element in Eq.~(\ref{Z(n+1)/2PQ}). So, the
continuous global symmetry of the model is
actually ${\rm U}(1)_{\rm PQ}\times
({\rm U}(1)_{\rm R}/Z_{(n+1)/2})$.

\par
As already explained, ${\rm U}(1)_{\rm R}$ is a
symmetry of the model only in the limit of
exact SUSY. In the scalar potential $V$ which
includes the soft SUSY-breaking terms too, it
is explicitly broken to its discrete subgroup
$Z_\mathcal{M}$. Thus, the group of
global symmetries of $V$ (except $Z^P_2$) is
${\rm U}(1)_{\rm PQ}\times (Z_{\mathcal{M}}/
Z_{(n+1)/2})$. The $Z_\mathcal{M}$ symmetry can
be factorized as follows:
\beq
Z_\mathcal{M}=Z_{n+3}\times Z_{(n+1)/2},
\label{ZMZ(n+3)Z(n+1)/2}
\eeq
where $Z_{n+3}$ is generated by
\beq
e^{i\frac{2\pi}{n+3}}\in {\rm U}(1)_{\rm R}.
\label{Z(n+3)}
\eeq
So, the global symmetry of $V$ takes the simple
form ${\rm U}(1)_{\rm PQ}\times Z_{n+3}\times
Z^P_2$.

\par
In any non-trivial minimum (actually, for any
fixed non-zero values of $P$, $Q$ and $S$),
this symmetry is spontaneously broken to
the $Z_2^{\rm mp}$ subgroup of
${\rm U}(1)_{\rm PQ}$. In case there was, after
inflation, a phase transition from the origin
in field space ($P=Q=S=0$) to a non-trivial
minimum, we would encounter copious production
of axionic strings \cite{axionwalls} as well as
domain walls. As we perform a full rotation
around such a string, the phase of $P$ or $Q$
changes, respectively, by $-2\pi$ or $2\pi$.
There are two types of walls separating vacua
which are related either by the group element
in Eq.~(\ref{Z(n+3)}) or the generator of
$Z_2^P$. Note that $n+3$ walls of the former
type can terminate together on the same line.

\par
As explained in the previous section, after the
onset of instantons at the QCD transition,
${\rm U}(1)_{\rm PQ}$ is explicitly broken to
$Z_{\mathcal{N}}$ with $\mathcal{N}=6(n+1)$
which is generated by the element
\beq
e^{i\frac{2\pi}{6(n+1)}}\in{\rm U}(1)_{\rm PQ}.
\label{ZN}
\eeq
So, after instantons, the axionic strings
become \cite{axionwalls} boundaries of $3(n+1)$
axionic walls \cite{sikivie} which separate
vacua related by the group element in
Eq.~(\ref{ZN}). We see that, if, after
inflation, a transition from
the origin in field space to a non-trivial
minimum takes place, a rich system of domain
walls is produced leading to a cosmological
catastrophe. So, it is clear that such a
transition should be avoided in our model.

\subsection{The shifted valley of minima}
\label{sec:shift}

\par
For $|S|\ll |P|\sim |Q|$, the F-terms in
Eqs.~(\ref{FP})-(\ref{FS}) can be approximated
as follows:
\bea
F_P&=&\frac{n+3}{2}\lambda_{\frac{n+3}{4}}
\frac{(PQ)^{\frac{n+1}{2}}Q}{M_{\rm P}^n}
\nonumber \\
& &+\frac{n-1}{2}\lambda_{\frac{n-1}{4}}
\frac{S^4(PQ)^{\frac{n-3}{2}}Q}{M_{\rm P}^n}
+\dots,
\label{FPapprox}
\eea
\bea
F_Q&=&\frac{n+3}{2}\lambda_{\frac{n+3}{4}}
\frac{(PQ)^{\frac{n+1}{2}}P}{M_{\rm P}^n}
\nonumber \\
& &+\frac{n-1}{2}\lambda_{\frac{n-1}{4}}
\frac{S^4(PQ)^{\frac{n-3}{2}}P}{M_{\rm P}^n}
+\dots,
\label{FQapprox}
\eea
\beq
F_S=4\lambda_{\frac{n-1}{4}}
\frac{S^3(PQ)^{\frac{n-1}{2}}}{M_{\rm P}^n}+
\dots,
\label{FSapprox}
\eeq
while the expression in the brackets in the RHS
of Eq.~(\ref{Vsoft}) takes the form
\beq
\left(A\lambda_{\frac{n+3}{4}}
\frac{(PQ)^{\frac{n+3}{2}}}{M_{\rm P}^n}
+A\lambda_{\frac{n-1}{4}}
\frac{S^4(PQ)^{\frac{n-1}{2}}}{M_{\rm P}^n}
+{\rm h.c.}\right)+\dots,
\label{Atermapprox}
\eeq
where the ellipses in these equations represent
terms of higher order in $S$. Note that the
second term in the RHS of Eqs.~(\ref{FPapprox})
and (\ref{FQapprox}) exists only for $n\geq 5$
($l\geq 1$), while the second term in
Eq.~(\ref{Atermapprox}) exists for all values
of $n$ in Eq.~(\ref{n4l1}). Using these
relations, we can expand, in this regime, the
potential $V$ in Eq.~(\ref{V}):
\bea
V&=&\frac{(n+3)^2}{4}
|\lambda_{\frac{n+3}{4}}|^2
\frac{|PQ|^{n+1}(|P|^2+|Q|^2)}{M_{\rm P}^{2n}}
\nonumber \\
& &+m_P^2|P|^2+m_Q^2|Q|^2-2|A|
|\lambda_{\frac{n+3}{4}}|
\frac{(|P||Q|)^{\frac{n+3}{2}}}{M_{\rm P}^n}
\nonumber \\
& &\times\cos\left[\frac{n+3}{2}
(\theta_P+\theta_Q)\right]+\dots
\nonumber \\
&\equiv& V_{(0)}+\dots,
\label{Vapprox}
\eea
where $\theta_P$ and $\theta_Q$ are the
phases of the complex scalar fields $P$ and $Q$
respectively. Here we assumed, without loss of
generality, that the product
$A\lambda_{\frac{n+3}{4}}$ is real and
negative, which can be readily achieved by
redefining the phase of the product of fields
$PQ$.

\par
The leading order part $V_{(0)}$ of the
potential
$V$ in Eq.~(\ref{Vapprox}) (consisting of the
explicitly displayed terms in the RHS of this
equation) is minimized with respect to the
phases $\theta_P$ and $\theta_Q$ for
\beq
\frac{n+3}{2}(\theta_P+\theta_Q)=0\quad
{\rm modulo}\quad 2\pi.
\label{mod1}
\eeq
Under this restriction on the phases, the
extrema of $V_{(0)}$ with respect to $|P|$ and
$|Q|$ are given by the conditions
\bea
\frac{\partial V_{(0)}}{\partial|P|}&=&
\frac{(n+3)^2}{4}|\lambda_{\frac{n+3}{4}}|^2
[(n+3)|P|^2+(n+1)|Q|^2]
\nonumber \\
& &\times\frac{|P|^n|Q|^{n+1}}{M_{\rm P}^{2n}}+
2m_P^2|P|-2|A|\frac{(n+3)}{2}
|\lambda_{\frac{n+3}{4}}|
\nonumber \\
& &\times
\frac{|P|^{\frac{n+1}{2}}|Q|^{\frac{n+3}{2}}}
{M_{\rm P}^n}=0
\label{Pderiv}
\eea
and
\bea
\frac{\partial V_{(0)}}{\partial|Q|}&=&
\frac{(n+3)^2}{4}|\lambda_{\frac{n+3}{4}}|^2
[(n+1)|P|^2+(n+3)|Q|^2]
\nonumber \\
& &\times\frac{|P|^{n+1}|Q|^n}{M_{\rm P}^{2n}}+
2m_Q^2|Q|-2|A|\frac{(n+3)}{2}
|\lambda_{\frac{n+3}{4}}|
\nonumber \\
& &\times
\frac{|P|^{\frac{n+3}{2}}|Q|^{\frac{n+1}{2}}}
{M_{\rm P}^n}=0.
\label{Qderiv}
\eea
Multiplying Eqs.~(\ref{Pderiv}) and
(\ref{Qderiv}) by $|P|$ and $|Q|$ respectively
and subtracting, we obtain
\bea
& &\frac{(n+3)^2}{4}|\lambda_{\frac{n+3}{4}}|^2
(|P|^2-|Q|^2)\frac{|P|^{n+1}|Q|^{n+1}}
{M_{\rm P}^{2n}}
\nonumber \\
& &+(m_P^2|P|^2-m_Q^2|Q|^2)=0,
\label{subtract}
\eea
which, for $m_P^2=m_Q^2\equiv m^2>0$ (see
Eq.~(\ref{mPmQm})), implies that $|P|=|Q|$.
Substituting $|P|$ for $|Q|$ in any of the
Eqs.~(\ref{Pderiv}) and (\ref{Qderiv}), we then
obtain $|P|$=0 or
\bea
& &\frac{(n+3)^2}{4}|\lambda_{\frac{n+3}{4}}|^2
(2n+4)\frac{|P|^{2(n+1)}}{M_{\rm P}^{2n}}+
2m^2
\nonumber \\
& &-2|A|\frac{(n+3)}{2}
|\lambda_{\frac{n+3}{4}}|
\frac{|P|^{n+1}}{M_{\rm P}^n}=0.
\label{min}
\eea
This equation has two real and positive
solutions for $$|A|^2>4(n+2)m^2,$$ which are
given by
\beq
\left(\frac{|P|^{n+1}}{M_{\rm P}^n}\right)_\pm
\equiv x_\pm=
\frac{|A|\pm\sqrt{|A|^2-4(n+2)m^2}}
{(n+2)(n+3)|\lambda_{\frac{n+3}{4}}|}.
\label{xpm}
\eeq
Obviously, $x=0$ and $x=x_+$ correspond to
(local) minima of $V_{(0)}$, while $x=x_-$
corresponds to a local maximum (here $x\equiv
|P|^{n+1}/M_{\rm P}^n$). So, $V_{(0)}$ has a
trivial minimum at $|P|=|Q|=0$ and a
``shifted'' minimum at $|P|=|Q|\sim
(m_{3/2}M_{\rm P}^n)^{\frac{1}{n+1}}$, which is
the absolute minimum for $|A|>(n+3)m$. Note
that the presence of the term
$\lambda_{\frac{n+3}{4}}(PQ)^{\frac{n+3}{2}}/
M_{\rm P}^n$ in the superpotential of
Eq.~(\ref{W}) is vital to the existence of the
shifted minimum. In view of the $Z_2^P$
symmetry, however, this superpotential term can
only exist if $(n+3)/2$ is an even positive
integer, which implies the restriction in
Eq.~(\ref{n4l1}).

\par
The trivial minimum of $V_{(0)}$ cannot be
consistent with our starting hypothesis that
$|S|\ll |P|\sim |Q|$ and should, thus, be
discarded. However, as we have already shown,
it happens to exist as a minimum of the full
potential $V$ in Eq.~(\ref{V}) with respect
to $|P|$ and $|Q|$ for all values of $|S|$ and
constitutes the trivial valley of minima.

\par
To see how the shifted minimum of $V_{(0)}$
evolves as $|S|$ increases from zero, we
consider the dominant $S$-dependent part of the
potential $V$ for $|S|\ll |P|\sim|Q|$, which is
given by
\bea
V_{(1)}&=&m_S^2|S|^2+
\Big[A\lambda_{\frac{n-1}{4}}
\frac{S^4(PQ)^{\frac{n-1}{2}}}{M_{\rm P}^n}
\nonumber \\
& &+
\frac{(n-1)(n+3)}{4}
\lambda_{\frac{n-1}{4}}
\lambda_{\frac{n+3}{4}}^*
S^4(P^*Q^*)^2
\nonumber \\
& &\times(|P|^2+|Q|^2)\frac{(|P||Q|)^{n-3}}
{M_{\rm P}^{2n}}+{\rm h.c.}\Big],
\label{V1}
\eea
where the first term in the brackets
corresponds to the second term in
Eq.~(\ref{Atermapprox}), while the second term
in the brackets originates from the
interference of the two explicitly displayed
terms in the RHS of Eq.~(\ref{FPapprox}) plus
the interference of the two explicitly
displayed terms in the RHS of
Eq.~(\ref{FQapprox}). So, the first term in the
brackets exists for all values of $n$, while
the second only for $n\geq 5$. Note that
$V_{(1)}$ contains the next-to-leading and the
next-to-next-to-leading order parts of $V$ in
the expansion of Eq.~(\ref{Vapprox}), which are
quadratic and quartic in $S$ respectively.
Actually, for $|P|$ and $|Q|$ at the
shifted minimum of $V_{(0)}$, all the terms
in $V_{(0)}$ are of the same order of magnitude,
while the first (mass) term in the RHS of
Eq.~(\ref{V1}) is suppressed by $(|S|/|P|)^2$
and the terms in the brackets by $(|S|/|P|)^4$.

\par
The minimization conditions for $V$ with
respect to $|P|$ and $|Q|$ coincide to leading
order with Eqs.~(\ref{Pderiv}) and
(\ref{Qderiv}). The dominant corrections to
these conditions for $|S|\ll |P|\sim |Q|$
originate from the next-to-next-to-leading
order terms in $V_{(1)}$ and are, thus,
suppressed
by $(|S|/|P|)^4$. So, for $|S|\ll |P|\sim |Q|$,
the shifted minimum of $V_{(0)}$ is also a
minimum of $V$ with respect to $|P|$ and $|Q|$
at a practically $S$-independent position. As a
consequence, for small values of $|S|$, we
obtain a ``shifted'' valley of minima of $V$ at
almost constant values of $|P|$ and $|Q|$. This
valley has obviously negative inclination for
non-zero and small values of $|S|$, due to the
negative mass term of $S$. It starts from the
``shifted'' saddle point of $V$ which lies at
$|S|=0$ and $|P|$, $|Q|$ equal to their values
at the shifted minimum of $V_{(0)}$. Let us
note, in passing, that a shifted valley of
minima was first used in Ref.~\cite{shift} as
an inflationary trajectory in order to avoid
the overproduction of doubly charged
\cite{magg} magnetic monopoles at the end of
hybrid inflation in a SUSY Pati-Salam \cite{ps}
GUT model.

\subsection{The PNGB curvaton}
\label{sec:pngb}

\par
The dominant $S$-dependent part of $V$ can be
expressed in terms of the phases of $P$, $Q$
and $S$ as follows:
\bea
V_{(1)}&=&m_S^2|S|^2-2|A|
|\lambda_{\frac{n-1}{4}}|
\frac{|S|^4(|P||Q|)^{\frac{n-1}{2}}}
{M_{\rm P}^n}
\nonumber \\
& &\times\cos\left(4\theta_S+\frac{n-1}{2}
(\theta_P+\theta_Q)\right)
\nonumber \\
& &+\frac{(n-1)(n+3)}{2}|
\lambda_{\frac{n-1}{4}}|
|\lambda_{\frac{n+3}{4}}||S|^4(|P|^2+|Q|^2)
\nonumber \\
& &\times\frac{(|P||Q|)^{n-1}}
{M_{\rm P}^{2n}}\cos\left(4\theta_S-2
(\theta_P+\theta_Q)\right),
\label{V1angles}
\eea
where we assumed that
$A\lambda_{\frac{n-1}{4}}$ is real and negative,
which can be readily arranged by redefining the
phase of $S$. Note that all the other
$A\lambda_k$'s (except
$A\lambda_{\frac{n+3}{4}}$ and
$A\lambda_{\frac{n-1}{4}}$) remain in general
complex since there is no extra field-rephasing
freedom left. From the preceding discussion, we
see that, on the shifted valley of minima
of $V$, the second and third term in the RHS of
Eq.~(\ref{V1angles}) are of the same order of
magnitude. For $n=1$, the third term vanishes.
Moreover, using Eq.~(\ref{xpm}), one can show
that, for $n\geq 5$, the coefficient of the
cosine in the second term is always greater in
absolute value than the coefficient of the
cosine in the third term. Under these
circumstances, $V_{(1)}$ is minimized with
respect to the phases $\theta_P$, $\theta_Q$,
$\theta_S$ of the fields by taking
\beq
4\theta_S+\frac{n-1}{2}(\theta_P+\theta_Q)=0
\quad{\rm modulo}\quad 2\pi.
\label{mod2}
\eeq
This together with Eq.~(\ref{mod1}) implies
that
\beq
4\theta_S-2(\theta_P+\theta_Q)=0\quad
{\rm modulo}\quad 2\pi,
\label{mod3}
\eeq
which maximizes the third term in the RHS of
Eq.~(\ref{V1angles}).

\par
We can define the real canonically normalized
fields corresponding to the phases of $P$, $Q$,
$S$ as follows:
\beq
\phi_P\equiv\sqrt{2}|P|\theta_P,\quad
\phi_Q\equiv\sqrt{2}|Q|\theta_Q,\quad
\phi_S\equiv\sqrt{2}|S|\theta_S.
\label{phiPQS}
\eeq
On the shifted valley, the last term in the
RHS of Eq.~(\ref{Vapprox}) generates a
``mass-squared'' for the real canonically
normalized field
$\phi_{PQ}\equiv (\phi_P+\phi_Q)/\sqrt{2}$
given by
\beq
m_{PQ}^2=|A|\frac{(n+3)^2}{2}
|\lambda_{\frac{n+3}{4}}|x_+,
\label{mPQ}
\eeq
where $x_+\equiv (|P|^{n+1}/M_{\rm P}^n)_+$ is
given by Eq.~(\ref{xpm}) and $m_{PQ}^2$ is of
order $m^2_{3/2}$. Also, the second and third
term in the RHS of Eq.~(\ref{V1angles})
generate on this valley ``masses-squared'',
respectively, for the combinations
\bea
\phi_{S1}&=&\frac{4\sqrt{2}|P|\phi_S+(n-1)|S|
\phi_{PQ}}{\sqrt{32|P|^2+(n-1)^2|S|^2}},
\nonumber \\
\phi_{S2}&=&\frac{\sqrt{2}|P|\phi_S
-|S|\phi_{PQ}}{\sqrt{2|P|^2+|S|^2}}
\label{phiS1S2}
\eea
given, respectively, by
\bea
m^2_{S1}&=&16|A||\lambda_{\frac{n-1}{4}}|x_+
\frac{|S|^2}{|Q|^2}\left(1+ \frac{(n-1)^2}{32}
\frac{|S|^2}{|Q|^2}\right),
\nonumber \\
m^2_{S2}&=&-8(n-1)(n+3)
|\lambda_{\frac{n-1}{4}}|
|\lambda_{\frac{n+3}{4}}|x_+^2
\frac{|S|^2}{|Q|^2}
\nonumber \\
& &\times\left(1+\frac{|S|^2}{2|Q|^2}\right),
\label{mS12}
\eea
which are suppressed by $|S|^2/|Q|^2$ relative
to $m^2_{3/2}$. Note that the field
$a=(\phi_P-\phi_Q)/\sqrt{2}$ remains massless
to all orders in perturbation theory since it
corresponds to the axion. So, we obtain a
2-dimensional subspace of massive fields
spanned by $\phi_{PQ}$ and $\phi_S$ which are
orthogonal to the massless axion direction.
The quadratic form obtained by summing the
three ``mass terms'' corresponding to the
``masses-squared'' in Eqs.~(\ref{mPQ}) and
(\ref{mS12}) must then be diagonalized to find
the two mass eigenstates and eigenvalues. This
task is particularly simple in the limit
$|S|\ll |P|$. To lowest order in $|S|/|P|$, the
eigenstates coincide with $\phi_{PQ}$ and
$\phi_S$, which we will also call $\sigma$ as
it will be our PNGB curvaton (see below). The
masses-squared of these fields on the
shifted valley are equal to $m_{PQ}^2$ in
Eq.~(\ref{mPQ}) and
\bea
\tilde{m}^2_\sigma&=&\frac{8}{n+2}
|\lambda_{\frac{n-1}{4}}|x_+\frac{|S|^2}{|P|^2}
\Big[(n+5)|A|
\nonumber \\
& &-(n-1)\sqrt{|A|^2-4(n+2)m^2}\Big]
\label{massS}
\eea
respectively. Obviously, $\tilde{m}^2_\sigma$ is
positive and suppressed by $|S|^2/|P|^2$
relative to $m_{PQ}^2$, which is of order
$m_{3/2}^2$. So, the field $\phi_S\equiv\sigma$
is a light PNGB when the system rolls down the
shifted valley of the potential with
$|S|\ll|P|$,
while $\phi_{PQ}$ is a massive field. Inclusion
of higher order corrections in the potential
$V$ along the shifted valley does not
change this situation.

\par
The discrete $Z_2^P$ symmetry of the model,
under which $P$ changes sign, is very important
for the PNGB nature of $\phi_S$. Without this
symmetry, the next-to-leading order term in the
RHS of Eqs.~(\ref{FPapprox}) and
(\ref{FQapprox}) would be proportional to $S^2$
rather than $S^4$, the leading order term in
the RHS of Eq.~(\ref{FSapprox}) would be linear
in $S$ rather than cubic, and the
next-to-leading order term in
Eq.~(\ref{FSapprox}) would contain $S^2$
instead of $S^4$. As a consequence, the terms
in the dominant $S$-dependent part of $V$ for
$|S|\ll |P|\sim |Q|$ which depend on the
phases of the fields would be quadratic in $S$
(compare with Eqs.~(\ref{V1}) and
(\ref{V1angles})) and, thus, the mass of
$\phi_S$ on the shifted valley would be of
order $m_{3/2}$.

\subsection{Cosmological evolution}
\label{sec:cosmo}

\par
SUGRA corrections \cite{crisis,cllsw,randall}
during inflation will add to $A$ a term
proportional to the Hubble parameter during
inflation, which is of order $m_{3/2}$ in
our case. To simplify the discussion, we take
these corrections to be universal. Also, the
masses-squared $m^2_P$, $m^2_Q$ and $m^2_S$
will acquire corrections proportional to $H^2$.
We assume that these corrections are positive
and, for simplicity, we also take them
universal at least for $m^2_P$ and $m^2_Q$. So,
nothing changes in the above discussion and
formulae after including the SUGRA corrections
during inflation except that $A$, $m^2$ and
$m^2_S$ must now be replaced by their effective
values
\beq
\bar{A}=A+c_AH,\quad\bar{m}^2=m^2+c_{PQ}H^2,
\quad\bar{m}^2_S=m^2_S+c_SH^2
\label{effective}
\eeq
respectively. Here, $c_A$ is a complex
parameter of order unity, while $c_{PQ}$ and
$c_S$ are real and positive parameters again
of order unity.

\par
The effective parameters $\bar{A}$ and
$\bar{m}$ are of the same order of magnitude as
$A$ and $m$, i.e. they are of order $m_{3/2}$.
The mass-squared of $S$, which is taken
negative, receives positive corrections from
SUGRA, which are of the same order of magnitude
as $m^2_S$. We can arrange the parameters so
that the effective mass-squared of $S$ is
positive during the initial stages of
inflation. In this case, the
shifted saddle point of $V$ becomes a local
minimum of the effective potential and the
system may be initially trapped in this minimum
during inflation. As $H$ decreases gradually
during inflation, the SUGRA corrections become
smaller and, at some moment of time, this
minimum may turn into a saddle point. The
system then slowly rolls down the shifted
valley which has a very small slope given by
the small effective mass-squared of $S$
\cite{footnote4}. During this slow roll,
$\phi_S$ is an effectively massless PNGB which
can act as curvaton.

\par
After the end of inflation, the system keeps
rolling down the shifted valley and
eventually ends up in damped oscillations about
a non-trivial minimum of $V$ where all the
fields $P$, $Q$ and $S$ acquire non-zero values
of order $(m_{3/2}M_{\rm P}^n)^{\frac{1}{n+1}}$
and the PQ symmetry is broken. Note that, if,
in the initial stages of inflation, the system
happened to be trapped in the trivial saddle
point of the potential $V$ at $P=Q=S=0$ (which,
of course, is a local minimum of the effective
potential in this case), it would later enter
into the trivial valley along the $S$ axis
rather than the shifted one. So, it would
end up in the minimum of Eq.~(\ref{trmin}),
where $P=Q=0$. This is obviously highly
undesirable since, in this case, the PQ
symmetry remains unbroken and no $\mu$ term is
generated. It is also important that, in our
case, the fields $P$, $Q$ and $S$ have non-zero
values during inflation after the time when the
cosmological scales exit the horizon as they
lie on the shifted valley. So, the global
symmetry of the model is already broken to
$Z_2^{\rm mp}$ during the relevant part of
inflation. Consequently, neither domain walls
nor axionic strings \cite{axionwalls} are
generated as the system settles in a
non-trivial minimum of $V$. Also, no axionic
walls \cite{sikivie} appear at the QCD
transition since the spontaneous breaking of
$Z_{\mathcal{N}}$ to $Z_2^{\rm mp}$ takes place
before the relevant part of inflation.
Therefore, no cosmological catastrophe is
encountered.

\section{Curvaton physics}
\label{sec:curvphys}

\subsection{The curvaton potential}
\label{sec:curvpot}

\par
From Eqs.~(\ref{V}), (\ref{FSapprox}) and
(\ref{V1angles}), we find that the dominant
part of the scalar potential which is relevant
to our curvaton candidate, in the case when
\mbox{$|S|\ll |P|\sim |Q|$}, is given by
\bea
V_{\rm curv} & = & \bar m_S^2|S|^2-2|\bar A|
|\lambda_{\frac{n-1}{4}}|
\frac{|S|^4(|P||Q|)^{\frac{n-1}{2}}}
{M_{\rm P}^n}
\nonumber \\
& &\times\cos\left[4\theta_S+\frac{n-1}{2}
(\theta_P+\theta_Q)\right]
\nonumber \\
& &+\frac{(n-1)(n+3)}{2}|
\lambda_{\frac{n-1}{4}}|
|\lambda_{\frac{n+3}{4}}||S|^4(|P|^2+|Q|^2)
\nonumber \\
& &\times\frac{(|P||Q|)^{n-1}}
{M_{\rm P}^{2n}}\cos\left[4\theta_S-2
(\theta_P+\theta_Q)\right]\nonumber \\
& & +16|\lambda_{\frac{n-1}{4}}|^2
\frac{(|P||Q|)^{n-1}}{M_{\rm P}^{2n}}|S|^6
+\cdots\,,
\label{Vcurv}
\eea
where we included the SUGRA corrections during
inflation and the ellipsis denotes terms of
higher order in $|S|$, which are,
therefore, subdominant. In the above, according
to Eq.~(\ref{effective}), we have
\beq
\bar m^2_S=c_SH^2-|m^2_S|\,,
\label{mS}
\eeq
where \mbox{$c_S\sim +1$} and
\mbox{$|m^2_S|\sim m_{3/2}^2$}.

\par
The potential $V_{\rm curv}$ is simplified by
considering that the $|P|$ and $|Q|$ fields have
already assumed their minimum value on the
shifted valley as given by Eq.~(\ref{xpm})
with $A$ and $m$ replaced by $\bar{A}$ and
$\bar{m}$ respectively. Furthermore, we can set
the phases $\theta_P$ and $\theta_Q$ equal to
zero since we are only interested in the
curvaton field direction, which practically
corresponds to $\theta_S$ for $|S|\ll|P|$.
Then, in view also of Eq.~(\ref{phiPQS}), the
above potential becomes
\bea
V_{\rm curv} & \simeq & \bar m_S^2|S|^2-\kappa
\frac{|\bar{A}|^2|S|^4}{|P|_{\rm val}^2}\cos
\left(2\sqrt{2}\frac{\phi_S}{|S|}\right)
\nonumber\\
& & +16|\lambda_{\frac{n-1}{4}}|^2
\frac{|P|_{\rm val}^{2(n-1)}}
{M_{\rm P}^{2n}}|S|^6+\cdots,
\label{Vcurv2}
\eea
where the value $|P|_{\rm val}$ of $|P|$ on the
shifted valley is given by
\mbox{$|P|_{\rm val}^{n+1}/M_{\rm P}^n\equiv
x_{+}$} (with $x_{+}$ from Eq.~(\ref{xpm}),
where $A$ and $m$ are replaced by $\bar{A}$ and
$\bar{m}$ respectively) and
\beq
\kappa\equiv
\left|\frac{\lambda_{\frac{n-1}{4}}}
{\lambda_{\frac{n+3}{4}}}\right|
\frac{(1+Z)
\left[(n+5)-(n-1)Z\right]}{(n+2)^2(n+3)}
\label{lambda}
\eeq
with $Z$ being
\beq
Z\equiv
\sqrt{1-4(n+2)\left(\frac{\bar{m}}{|\bar{A}|}
\right)^2}.
\label{Z}
\eeq
Note that $|P|_{\rm val}$ is practically
constant when the cosmological scales exit the
inflationary horizon since the shifted
valley is almost $|S|$-independent for
$|S|\ll|P|$ as shown in Sec.~\ref{sec:shift}
and the Hubble parameter is very slowly varying
(see Eq.~(\ref{vareps})).

\par
From Eq.~(\ref{Z}), it is evident that
\mbox{$0<Z<1$} and, therefore, $\kappa$ is
positive \cite{footnote5}.
Hence, setting
\beq
\sigma\simeq\phi_S\quad{\rm and}\quad
v\simeq\frac{1}{2\sqrt 2}|S|,
\label{svs}
\eeq
we obtain the curvaton potential for
\mbox{$v\ll v_0$} (the value of $v$ in the
vacuum) as
\beq
V(\sigma)\simeq 64\kappa|\bar{A}|^2\frac{v^4}
{|P|_{\rm val}^2}\left[1-\cos\left(
\frac{\sigma}{v}\right)\right].
\eeq
Considering that the value of $v_0$ is given by
the value $|S|_0$ of $|S|$ in the vacuum, for
which \mbox{$|S|_0\sim |P|_{\rm val}$}, we find
\beq
V(\sigma)\sim\kappa|\bar{A}|^2\frac{v^4}
{v_0^2}\left[1-\cos\left(\frac{\sigma}{v}
\right)\right],
\label{vcurv}
\eeq
where [cf. Eq.~(\ref{eq:inter})]
\beq
v_0\sim M_{\rm I}\sim
(m_{3/2}M_{\rm P}^n)^{\frac{1}{n+1}}.
\label{fPQ}
\eeq
Comparing the above with Eq.~(\ref{Vs}), we see
that the mass of the curvaton is given by
\beq
\tilde m_\sigma\sim\sqrt{\kappa}\,|\bar{A}|
\left(\frac{v}{v_0}\right),
\eeq
which agrees with Eq.~(\ref{massS}). Therefore,
when the cosmological scales exit the horizon,
we have
\beq
\tilde m_\sigma\sim\varepsilon m_{3/2},
\eeq
where we used Eq.~(\ref{eps}) and considered
that \mbox{$|\bar{A}|\sim m_{3/2}$} and
\mbox{$\kappa\sim 1$}. Hence, because, in the
modular inflation model that we are
considering, we have \mbox{$H_*\sim m_{3/2}$},
we find that, since \mbox{$\varepsilon\ll 1$},
$\sigma$ is, as required, effectively massless
when the cosmological scales exit the horizon
during inflation.

\par
The order parameter $v$ for our curvaton field
is determined by the value of $|S|$, for which
the potential in Eq.~(\ref{Vcurv2}), when
taking $\sigma\rightarrow 0$, becomes
\beq
V(|S|)\simeq\bar m_S^2|S|^2-\kappa
\frac{|\bar{A}|^2|S|^4}{|P|_{\rm val}^2}+
\kappa_S^2|\bar{A}|^2\frac{|S|^6}
{|P|_{\rm val}^4}+\cdots,
\label{|S|}
\eeq
where
\beq
\kappa_S\equiv\left
|\frac{\lambda_{\frac{n-1}{4}}}
{\lambda_{\frac{n+3}{4}}}\right|
\frac{4(1+Z)}{(n+2)(n+3)}
\label{lambdaS}
\eeq
and we have used Eq.~(\ref{xpm}). The minimum
of the above potential occurs at
\beq
|S|\simeq\frac{1}{\kappa_S}
\sqrt{\frac{2\kappa}{3}}\,
|P|_{\rm val}\sim M_{\rm I},
\label{S0}
\eeq
which, after the end of inflation, becomes
equal to $|S|_0$ as $\bar{A}$ and $\bar{m}$
are replaced by $A$ and $m$ respectively.

\subsection{\boldmath The required
$\varepsilon$}
\label{sec:varepsilon}

\par
Let us now calculate the value of $\varepsilon$
required so that our curvaton scenario works.
Firstly, we note that, in our case, the
curvaton assumes a random value at the phase
transition at which, during inflation, the
system leaves the shifted saddle point of
the potential and starts slowly rolling down
the shifted valley (see
Sec.~\ref{sec:cosmo}). This value typically is
\mbox{$\sigma\sim v$}. After the end of
inflation and before the onset of the
oscillations, the phase $\theta$ corresponding
to the curvaton degree of freedom is overdamped
and remains frozen. Hence, we expect that, at
the onset of the oscillations, we have
\begin{equation}
\sigma_{\rm osc}\sim\theta v_0,
\label{soscv0}
\end{equation}
where, typically,
\mbox{$\theta\simeq\theta_S\sim 1$} and
we took into account that the order parameter
assumes its vacuum value very
soon after the end of inflation.
Combining Eqs.~(\ref{sosc}) and (\ref{soscv0}),
we find
\begin{equation}
\varepsilon\sim
\frac{\Omega_{\rm dec}}{\pi\zeta\theta}
\left(\frac{m_{3/2}}{M_{\rm P}}
\right)^{\frac{n}{n+1}},
\label{epszeta}
\end{equation}
where we also used Eq.~(\ref{fPQ}) and that
\mbox{$H_*\sim m_{3/2}$}. The $\varepsilon$
above is always larger than
$\varepsilon_{\rm min}$, where
\begin{equation}
\varepsilon_{\rm min}\sim
\left(\frac{m_{3/2}}{M_{\rm P}}
\right)^{\frac{n}{n+1}},
\label{epsmin}
\end{equation}
which is derived from Eq.~(\ref{epsbound}) with
\mbox{$H_*\sim m_{3/2}$}.

\par
Let us now enforce the constraint in
Eq.~(\ref{H*bound}), which, for
\mbox{$H_*\sim m_{3/2}$}, reads
\begin{equation}
\varepsilon<
\frac{\Omega_{\rm dec}^{\frac{1}{2}}}{\pi\zeta}
\left(\frac{M_{\rm P}}{T_{\rm BBN}}
\right)^{\frac{1}{2}}
\left(\frac{m_{3/2}}{M_{\rm P}}
\right)^{\frac{5}{4}}\sim
10^{-4}\Omega_{\rm dec}^{\frac{1}{2}}.
\label{epsbound0}
\end{equation}
From Eqs.~(\ref{epszeta}) and (\ref{epsbound0}),
it is easy to find that the above bound can be
satisfied only if $n$ is large enough:
\begin{equation}
n>\frac{8+\log(\Omega_{\rm dec}^{\frac{1}{2}}/
\theta)}{7-\log(\Omega_{\rm dec}^{\frac{1}{2}}/
\theta)}.
\label{nbound}
\end{equation}
In view of Eq.~(\ref{wmap}), we see that, for
\mbox{$\theta\sim 1$}, we have
\mbox{$n\geq 1$}.

\par
An upper bound on $n$ can be obtained by
requiring that the curvaton decays before BBN.
The interaction of $\sigma$ with ordinary
particles is governed by the effective $\mu$
term in Eq.~(\ref{eq:muterm}), which results
\cite{chun} into the following decay rate of
$\sigma$ into two Higgs particles:
\begin{equation}
\Gamma_\sigma\sim\frac{m_\sigma^3}{v_0^2}.
\label{Gs}
\end{equation}
Demanding that
\mbox{$\Gamma_\sigma\geq H_{\rm BBN}$} results
in the bound
\bea
& &\Gamma_\sigma\sim
10^{-\frac{30n}{n+1}}\left(
\frac{m_\sigma}{\rm TeV}\right)^3{\rm TeV}
\geq H_{\rm BBN}\sim 10^{-27}~{\rm TeV}
\nonumber\\
& &\Rightarrow\;
m_\sigma\gsim 10^{\frac{n-9}{n+1}}~{\rm TeV},
\label{nboundup}
\eea
where we used Eq.~(\ref{fPQ}). The above
requirement is always satisfied if
\mbox{$m_\sigma\gsim 10~{\rm TeV}$}. However,
in the opposite case where
\mbox{$m_\sigma<10~{\rm TeV}$}, we see
that it yields an upper bound on $n$:
\begin{equation}
n\leq\frac{9+\log(m_\sigma/{\rm TeV})}
{1-\log(m_\sigma/{\rm TeV})},
\label{nup}
\end{equation}
which, roughly, demands \mbox{$n\leq 9$} for
\mbox{$m_\sigma\lsim 1~{\rm TeV}$}.

\subsection{The reheating of the universe}
\label{sec:reheat}

\par
We will now proceed further by first discussing
the reheating of the universe. This requires
that we consider separately the cases when the
curvaton decays before or after it dominates
the universe.

\subsubsection{Curvaton decay before domination
($\Omega_{\rm dec}\ll 1$)}
\label{sec:before}

\par
During the radiation era and after the onset
of curvaton oscillations, for the curvaton
energy density fraction, we have
\mbox{$\rho_\sigma/\rho\propto a(t)\propto
H^{-1/2}$}, where $a(t)$ is the scale factor of
the universe. Hence, in this case, we find that
\begin{equation}
\Omega_{\rm dec}\sim
\left(\frac{\min\{m_\sigma,\Gamma_{\rm inf}\}}
{\Gamma_\sigma}\right)^{\frac{1}{2}}\left(
\frac{\sigma_{\rm osc}}{M_{\rm P}}\right)^2,
\label{bla}
\end{equation}
where we have used Eq.~(\ref{sosc}) and
\begin{equation}
\left.\frac{\rho_\sigma}{\rho}\right|_{\rm osc}
\sim\left(\frac{\sigma_{\rm osc}}{M_{\rm P}}
\right)^2,
\label{rhofracosc}
\end{equation}
which is derived from the fact that
\mbox{$\rho_\sigma|_{\rm osc}\simeq
\frac{1}{2}m_\sigma^2\sigma_{\rm osc}^2$} and
\mbox{$\rho_{\rm osc}\simeq
m_\sigma^2M_{\rm P}^2$}. Using Eq.~(\ref{Gs})
into Eq.~(\ref{bla}) and also Eqs.~(\ref{sosc})
and (\ref{fPQ}), we obtain
\begin{equation}
\varepsilon\sim\frac{g^{\frac{1}{2}}
\Omega_{\rm dec}^{\frac{1}{2}}}{\pi\zeta}
\left(\frac{m_{3/2}}{M_{\rm P}}\right)
^{\frac{1}{2}(\frac{n+2}{n+1})}.
\label{epszeta1}
\end{equation}
Here, we have also used that
\mbox{$\Gamma_{\rm inf}<H_*\sim
m_{3/2}\sim m_\sigma$} and
\begin{equation}
\Gamma_{\rm inf}\sim g^2m_{3/2}
\label{Ginf}
\end{equation}
with $g$ being the dimensionless coupling
constant of the inflaton to its decay products
and the mass of the inflaton field $s$ taken to
be \mbox{$m_s\lsim H_*\sim m_{3/2}$}.

\par
In principle, $g$ can be as low as
$m_s/M_{\rm P}$ if the inflaton decays
gravitationally. However, since reheating has
to occur before BBN, $g$ has to lie in the
range:
\begin{equation}
10^{-14}\sim 10\frac{m_{3/2}}{M_{\rm P}}<g<1,
\label{grange}
\end{equation}
where we used the fact that the reheat
temperature $T_{\rm reh}$, in this case, is
\beq
T_{\rm reh}\sim\sqrt{\Gamma_{\rm inf}M_{\rm P}}
\sim g\sqrt{m_{3/2}M_{\rm P}}.
\label{Treh}
\eeq
Combining Eqs.~(\ref{epszeta}) and
(\ref{epszeta1}), we find the relation
\begin{equation}
\frac{g}{\Omega_{\rm dec}}\sim\frac{1}{\theta^2}
\left(\frac{m_{3/2}}{M_{\rm P}}\right)
^{\frac{n-2}{n+1}},
\label{g1}
\end{equation}
which results in
\begin{equation}
n\simeq\frac{30-\log g
+2\log(\Omega_{\rm dec}^{\frac{1}{2}}/\theta)}
{15+\log g-2\log(\Omega_{\rm dec}^{\frac{1}{2}}
/\theta)}.
\label{n1}
\end{equation}
In view of Eqs.~(\ref{wmap}) and (\ref{grange}),
we see that, for \mbox{$\theta\sim 1$},
the allowed range for $n$ is
\begin{equation}
2\leq n\leq 44.
\label{nrange1}
\end{equation}
The lower bound in the above is tighter than
the bound in
Eq.~(\ref{nbound}). In fact, comparing
Eqs.~(\ref{nbound}) and (\ref{n1}), it
is easy to obtain the bound
\begin{equation}
g\Omega_{\rm dec}^{\frac{1}{2}}\leq 10^6\,
\theta,
\label{gW}
\end{equation}
which is easily satisfied provided that the
angle $\theta$ is not extremely small.

\subsubsection{Curvaton decay after
domination ($\Omega_{\rm dec}\approx 1$)}
\label{sec:after}

\par
In this case, the curvaton dominates the
energy density of the universe when
\mbox{$H=H_{\rm dom}$}, where $H_{\rm dom}$ is
given by
\begin{equation}
H_{\rm dom}\sim
\left(\frac{\sigma_{\rm osc}}{M_{\rm P}}
\right)^4\min\{m_\sigma, \Gamma_{\rm inf}\}.
\label{Hdom}
\end{equation}
Now, using Eqs.~(\ref{soscv0}), (\ref{Gs}) and
(\ref{Ginf}), it can be shown that the
requirement \mbox{$\Gamma_\sigma<H_{\rm dom}$}
results in the bound
\begin{equation}
g>\frac{1}{\theta^2}
\left(\frac{m_{3/2}}{M_{\rm P}}\right)
^{\frac{n-2}{n+1}},
\label{gW2}
\end{equation}
which yields
\begin{equation}
n>\frac{30-\log\,g-2\log\theta}
{15+\log\,g+2\log\theta}.
\label{n2}
\end{equation}
This provides a lower bound on $g$ for given
$n$ and $\theta$, reminiscent of Eq.~(\ref{n1})
with \mbox{$\Omega_{\rm dec}\approx 1$}. For
\mbox{$\theta\sim 1$}, the above bound implies
\beq
n\geq 2.
\label{nrange2}
\eeq
This time the hot big bang begins after the
decay of the curvaton, which suggests that the
reheat temperature is now given by
\beq
T_{\rm REH}\sim\sqrt{\Gamma_\sigma M_{\rm P}}
\sim m_{3/2}\left(\frac{m_{3/2}}{M_{\rm P}}
\right)^{\frac{1}{2}(\frac{n-1}{n+1})}.
\label{Treh1}
\eeq
It can be easily checked that the above is
higher that $T_{\rm BBN}$ when \mbox{$n\leq 9$},
in agreement with Eq.~(\ref{nup}).

\subsection{Avoiding axion overproduction}
\label{sec:axion}

\par
There is a stringent upper bound on the PQ
scale originating from the requirement that the
generated axions do not overclose the universe.
The typical mass of the axion is \cite{turner}
\beq
m_{\rm a}\sim 10^{-5}\left(
\frac{10^{12}~{\rm GeV}}{f_{\rm a}}\right)~
{\rm eV},
\eeq
where, for the PQ scale (axion decay constant),
we have \mbox{$f_{\rm a}\approx v_0$}
\cite{footnote6}. The above mass implies that
the onset of axion oscillations takes place
when the energy density of the universe is
\beq
\rho^{\frac{1}{4}}_{\rm axosc}
\sim\sqrt{m_{\rm a}M_{\rm P}}\sim 10^2
\left(\frac{10^{12}~{\rm GeV}}{v_0}
\right)^{\frac{1}{2}}{\rm GeV},
\label{Ta}
\eeq
where the subscript ``axosc'' indicates the
time at which axion oscillations begin. In
contrast, the energy density of the universe
when the curvaton decays is
\beq
\rho_{\rm dec}^{\frac{1}{4}}\sim
\sqrt{\Gamma_\sigma M_{\rm P}}\sim
10\left(\frac{10^{12}~{\rm GeV}}{v_0}
\right){\rm GeV},
\eeq
where we have used Eq.~(\ref{Gs}) and also that
\mbox{$m_\sigma\sim m_{3/2}$}. Now, from
Eq.~(\ref{fPQ}), we have
\beq
v_0\geq\sqrt{m_{3/2}M_{\rm P}},
\eeq
which suggests that, in all cases,
\beq
\rho_{\rm dec}<\rho_{\rm axosc}.
\eeq
Hence, the axion oscillations always start
before the curvaton decays. Thus, at the onset
of the axion oscillations, the ratio of the
axion energy density to the energy density of
the universe is
\beq
\left.\frac{\rho_{\rm a}}{\rho}\right|
_{\rm axosc}
\sim\theta_{\rm a}^2
\left(\frac{v_0}{M_{\rm P}}\right)^2
\sim\theta_{\rm a}^2
\left(\frac{m_{3/2}}{M_{\rm P}}\right)
^{\frac{2}{n+1}},
\label{rhoaxosc}
\eeq
where we have considered that the amplitude of
the axion field at the onset of its
oscillations is \mbox{$\sim\theta_{\rm a}
f_{\rm a}$}
with $\theta_{\rm a}$ being the initial
misalignment angle.

\par
Let us, firstly, investigate the case when the
curvaton decays before it dominates the
universe. In this case, the axion oscillates
in a radiation background until the time
$t_{\rm eq}$ of equal matter and radiation
energy densities, denoted hereafter by the
subscript ``eq''. Since the energy density of
the oscillating axion scales like pressureless
matter with the expansion of the universe, we
have that, until $t_{\rm eq}$,
\beq
\frac{\rho_{\rm a}}{\rho}\propto a,
\label{rar}
\eeq
where the scale factor of the universe
\mbox{$a\propto 1/T$}. In view of the above,
the requirement that axions do not overclose
the universe translates into requiring that
\beq
\left.\frac{\rho_{\rm a}}{\rho}\right|_{\rm eq}
\leq 1,
\label{rhoeq}
\eeq
since, by definition,
\mbox{$\rho_{\rm eq}\sim\rho_m$}, where
$\rho_m$ is the energy density of matter.
From Eqs.~(\ref{rhoaxosc}), (\ref{rar}) and
(\ref{rhoeq}), we obtain the bound
\beq
\theta_{\rm a}\leq\left(
\frac{10^{12}~{\rm GeV}}
{v_0}\right)^{\frac{3}{4}},
\label{thaxbound}
\eeq
where we used that \mbox{$T_{\rm eq}\approx 2.8
\times 10^{-9}~{\rm GeV}$}. Hence, an initial
misalignment angle of order unity is possible
only if \mbox{$v_0\leq 10^{12}~{\rm GeV}$}. In
view of Eq.~(\ref{fPQ}), this, in turn, is
possible only if \mbox{$n\leq 1$}.

\par
Thus, if $n>1$ and we insist on \mbox{
$\theta_{\rm a}\sim 1$}, the only way that
axion overproduction can be avoided is by
considering that the curvaton {\em does}
dominate the universe before decaying. In this
case, reheating due to curvaton decay dilutes
the axion energy density because of a dramatic
production of entropy. The entropy ratio at
curvaton decay is easily estimated as
\bea
\frac{{\cal S}_{\rm after}}
{{\cal S}_{\rm before}} & \sim &
\left(\frac{\rho_\sigma}{\rho_\gamma}\right)
_{\rm REH}^{\frac{3}{4}}
\sim\left(\frac{H_{\rm dom}}{\Gamma_\sigma}
\right)^{\frac{1}{2}}
\nonumber\\
& \sim & \frac{g\,\theta^2v_0^3}
{m_{3/2}M_{\rm P}^2}\sim
g\theta^2\left(\frac{M_{\rm P}}{m_{3/2}}
\right)^{\frac{n-2}{n+1}},
\label{SS}
\eea
where the subscript ``REH'' denotes the
time of the curvaton decay, $\rho_\gamma$ is
the energy density of the background radiation
due to the decay of the inflaton, and we have
used Eqs.~(\ref{fPQ}), (\ref{Gs}), (\ref{Ginf})
and (\ref{Hdom}) considering also that
\mbox{$\sigma_{\rm osc}\sim \theta\, v_0$}.
The exact calculation multiplies \cite{STurner}
the result by a factor of $1.83\,g_\star^{1/4}
\sim 1$, where \mbox{$g_\star\sim 10-10^2$} is
the effective number of relativistic degrees of
freedom. As expected, the
entropy production depends on the value of the
coupling constant $g$. Hence, avoiding axion
overproduction, which could overclose the
universe, is expected to set another lower
bound on $g$, more stringent than the one in
Eq.~(\ref{gW2}) \cite{footnote7}.

\par
The fact that the case of curvaton domination
requires a larger value of $g$
[cf. Eq.~(\ref{gW2})] is to be expected because
larger $g$ implies that the inflaton decays
earlier and, therefore, the energy density
fraction $\rho_\sigma/\rho$ grows substantially,
allowing the curvaton to dominate the universe
before its decay. The higher $g$ is the more
dominant the curvaton will be at its decay and,
hence, the more diluted the axion energy
density will become after the decay of the
curvaton.

\par
A crucial further requirement for the dilution
of the axion energy density is
\cite{df,choi,lazaetc,super} that the entropy
release occurs after the onset of axion
oscillations. Hence, the requirement is
\beq
\rho_{\rm axosc}\gg T_{\rm REH}^4.
\label{axbound}
\eeq

\subsection{The evolution of the order
parameter}
\label{sec:evol}

Let us now concentrate on the evolution of the
order parameter $v$, which has to be such as to
achieve the required value for $\varepsilon$.
The order parameter is determined by the
rolling $|S|$. When the cosmological scales
exit the horizon, the field $|S|$ has to be
slowly rolling because we need the order
parameter to vary slowly enough not to
destabilize the approximate scale invariance of
the perturbation spectrum
[cf. Sec.~\ref{sec:ns}]. Therefore, the
Klein-Gordon equation for $|S|$ takes the form
\begin{equation}
3H|\dot{S}|+\bar m_S^2|S|\simeq 0.
\label{kg}
\end{equation}
Using Eq.~(\ref{mS}), the rate of growth of the
order parameter in this case can be easily
found to be
\begin{equation}
\frac{\dot v}{v}=
\frac{|\dot S|}{|S|}=\frac{1}{3}c_S
\left(\frac{|m_S^2|}{c_SH^2}-1\right)H.
\label{fdotf}
\end{equation}

\par
The amplification factor $\varepsilon^{-1}$ can
be found as follows. Using Eq.~(\ref{VN}), we
can write $|S|$ as a function of the number $N$
of the remaining e-foldings of inflation.
Starting from Eq.~(\ref{kg}) and after a little
algebra, we obtain
\beq
\frac{3}{c_S}\frac{d\ln|S|}{dN}=
\frac{e^{-2F_sN_{\rm x}}-e^{-2F_sN}}
{1-e^{-2F_sN}},
\label{dSN}
\eeq
where $N_{\rm x}$ corresponds to the phase
transition which changes the sign of
$\bar m_S^2$. Here, we used the fact that, by
definition,
\beq
|m_S^2|\equiv c_SH^2_{\rm x}\simeq
c_SH_{\rm m}^2(1-e^{-2F_sN_{\rm x}}),
\label{Hx}
\eeq
where \mbox{$H_{\rm x}\equiv H(N_{\rm x})$} and
\mbox{$H_{\rm m}=\sqrt{V_{\rm m}}/\sqrt{3}
M_{\rm P}$} with $V_{\rm m}$ being the scale of
the inflaton potential as given in
Eq.~(\ref{V0}). Integrating Eq.~(\ref{dSN}), we
get
\bea
\frac{6}{c_S}\ln
\left(\frac{|S|_*}{|S|_{\rm x}}\right)
& = & (1-e^{-2F_sN_{\rm x}})F_s^{-1}\ln
\left(\frac{e^{2F_sN_{\rm x}}-1}
{e^{2F_sN_*}-1}\right)
\nonumber\\
& &-2(N_{\rm x}-N_*),
\label{SN}
\eea
where \mbox{$|S|_*\equiv |S|(N_*)$} and
\mbox{$|S|_{\rm x}\equiv |S|(N_{\rm x})$}.

\par
Now, it is straightforward to check that, if
\mbox{$2F_sN_*\gg 1$}, the
RHS of the above equation tends to zero, which
yields \mbox{$|S|_*\approx|S|_{\rm x}$}.
This is also understood by observing that, in
this case, $H^2(N)\simeq
H_{\rm m}^2(1-e^{-2F_sN})\approx H_{\rm m}^2$,
which means that \mbox{$\delta(H^2)/H^2\approx
2F_s\delta Ne^{-2F_sN}\ll 1$}, where
$\delta(H^2)=H^2_{\rm x}-H^2_*$ and $\delta N=
N_{\rm x}-N_*>0$. Thus, in the e-folding
interval $\delta N$, the effective mass-squared
of $S$ hardly changes: \mbox{$\bar m_S^2(N_*)
\approx\bar m_S^2(N_{\rm x})\equiv 0$}, i.e.
the mass is very close to zero. This implies
that $|S|$ remains frozen and, thus,
\mbox{$|S|_*\approx|S|_{\rm x}$}.
Moreover, the displacement of $|S|$ from
the origin at the phase transition is
determined by its quantum fluctuations, which
means that
\beq
|S|_{\rm x}\sim\frac{H_{\rm x}}{2\pi}.
\label{SN0}
\eeq
For \mbox{$2F_sN_*\gg 1$}, $H_{\rm x}\approx
H_*~(\approx H_{\rm m})$ and, thus, $|S|_*
\sim H_*/2\pi$. So, under these circumstances,
we have, for the amplification factor,
\mbox{$\varepsilon=\varepsilon_{\rm min}$},
which is defined in Sec.~\ref{sec:factor}.

\par
It is easily seen that, generally,
\beq
\varepsilon=\frac{|S|_*}{|S|_0}\sim
\frac{|S|_*}{H_*}\frac{H_*}{v_0}\quad
\Rightarrow\quad |S|_*\sim\frac{\varepsilon}
{\varepsilon_{\rm min}}H_*.
\label{See}
\eeq
Note that, since the required $\varepsilon$ is
always much bigger than
$\varepsilon_{\rm min}$, as we saw from
Eqs.~(\ref{epszeta}) and (\ref{epsmin}),
$|S|_*\gg H_*$. Furthermore, it is obvious from
the above discussion that, for
\mbox{$\varepsilon>\varepsilon_{\rm min}$}, we
need to have
\beq
F_s\lsim \frac{1}{2N_{\rm x}}.
\eeq
Finally, when \mbox{$2F_sN_{\rm x}\ll 1$},
Eq.~(\ref{SN}) reduces to
\beq
\frac{3}{c_S}\ln
\left(\frac{|S|_*}{|S|_{\rm x}}\right)
\simeq N_{\rm x}\left[\ln\left(\frac{N_{\rm x}}
{N_*}\right)-1\right]+N_*.
\label{SNo}
\eeq

\par
In view of Eq.~(\ref{fdotf}), the requirement in
Eq.~(\ref{nsbound}) takes the form
\beq
\frac{c_S}{3}\,(H^2_{\rm x}-H^2_*)\ll
H^2_*,
\eeq
where we took into account Eq.~(\ref{Hx}). This
inequality can be recast as
\beq
\frac{c_S}{3}\,e^{-2F_sN_*}\left(
\frac{1-e^{-2F_s(N_{\rm x}-N_*)}}
{1-e^{-2F_sN_*}}\right)\ll 1.
\label{nnn}
\eeq
When \mbox{$2F_sN_{\rm x}\ll 1$}, the above
reduces to
\beq
\frac{c_S}{3}\,\frac{N_{\rm x}-N_*}{N_*}\ll 1.
\label{NNN}
\eeq
This means that, in this case, the cosmological
scales must exit the horizon not much later
than the phase transition which changes the
sign of $\bar m_S^2$.

\par
Another issue to be addressed concerns the
requirement that $|S|$ {\em does} slow roll
at the time when the cosmological scales exit
the horizon, in contrast to the case when
the slope of the potential is so small that
the motion of $|S|$ is dominated by its quantum
fluctuations. Indeed, since
\mbox{$|\bar m_S^2(H_*)|\ll H_*^2$}, $|S|$ is
effectively massless and, hence, it obtains a
superhorizon spectrum of perturbations of order
$H_*/2\pi$, much like $\sigma$. In order for
its quantum fluctuations not to dominate its
motion, $|S|$ has to be outside the quantum
diffusion zone. The condition for this to occur
is
\beq
\left|\frac{\partial V}{\partial|S|}\right|_*
=2|\bar m_S^2(H_*)||S|_*\geq H_*^3\,.
\eeq
Using Eqs.~(\ref{mS}) and (\ref{Hx}) and working
as before, the above constraint is recast as
\bea
\ln\left(\frac{|S|_*}{|S|_{\rm x}}\right)
& \geq & 2F_sN_*-\ln\left(\frac{c_S}{\pi}
\right)
\nonumber\\
& & +\ln\left(\frac{1-e^{-2F_sN_*}}
{1-e^{-2F_s(N_{\rm x}-N_*)}}\right)
\nonumber\\
& & +\frac{1}{2}\ln\left(\frac{1-e^{-2F_sN_*}}
{1-e^{-2F_sN_{\rm x}}}\right),
\label{SNbound}
\eea
where we have also used Eq.~(\ref{SN0}). If
\mbox{$2F_sN_{\rm x}\ll 1$}, the above
equation becomes
\beq
\ln\left(\frac{|S|_*}{|S|_{\rm x}}\right)
+\ln\left(\frac{c_S}{\pi}\right)+\ln\left[\frac
{N_{\rm x}^{\frac{1}{2}}(N_{\rm x}-N_*)}
{N_*^{\frac{3}{2}}}\right]\geq 0.
\label{SNbound0}
\eeq

\section{\boldmath  A Concrete example:
$n=5$ \& $\theta\sim 1$}
\label{sec:ex}

\par
From the bounds on $n$ in Eqs.~(\ref{nup}),
(\ref{nrange1}) and (\ref{nrange2}) and also in
view of Eq.~(\ref{n4l1}), we see that not many
choices for $n$ are allowed. In fact, we can
only accept the cases corresponding to
\mbox{$l=1,2$} (i.e. \mbox{$n=5,9$}) with
the latter choice being marginal as far as the
BBN constraint in Eq.~(\ref{nup}) is concerned.
Hence, to illustrate the above, we present an
example taking \mbox{$n=5$} (i.e. \mbox{$l=1$}
in Eq.~(\ref{n4l1})) and considering that the
orthogonal axion assumes a random value after
the phase transition, i.e.
\mbox{$\theta\sim 1$}.

\par
\begin{figure}[tb]
\centering
\includegraphics[width=\linewidth]{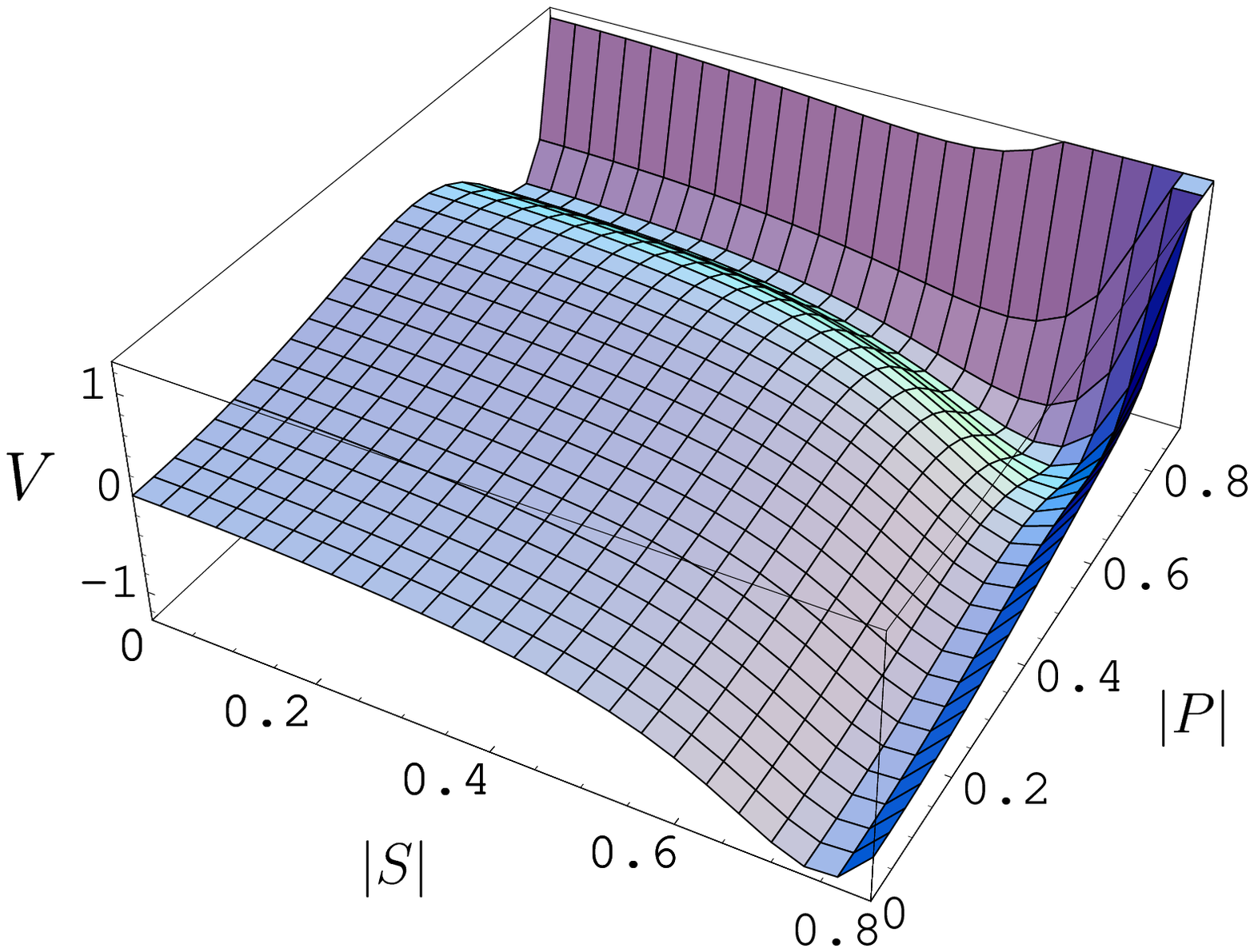}
\caption{Plot of the scalar potential
$V(|S|,|P|)$ which is defined in
Eqs.~(\ref{V})-(\ref{Vsoft}) in units of
$M_{\rm I}^{14}/M_{\rm P}^{10}$ with respect to
$|S|$ and $|P|$, which are measured in units of
$M_{\rm I}$. Here, we have taken \mbox{$n=5$}
and \mbox{$M_{\rm I}\equiv
(mM_{\rm P}^5)^{1/6}$}. We have also chosen
\mbox{$m_P^2=m_Q^2=-m_S^2\equiv m^2$},
\mbox{$A=-9\,m$}, \mbox{$\lambda_0=
\lambda_1=\lambda_2=1$}, \mbox{$|P|=|Q|$} and
\mbox{$\theta_S=\theta_P=\theta_Q=0$} so that
the potential is minimized according to
Eqs.~(\ref{mod1}) and (\ref{mod2}). The
shifted valley is clearly visible. It
starts at the value \mbox{$|P|\approx 0.814\,
M_{\rm I}$} when \mbox{$|S|=0$} and arcs down
towards the $|S|$ axis, passing through the
non-trivial minimum at \mbox{$|S|_0\approx
0.683\,M_{\rm I}$} and \mbox{$|P|_0\approx
0.772\,M_{\rm I}$}. The trivial
valley at \mbox{$|P|=0$} can also be discerned.}
\label{fig}
\end{figure}

The bound in Eq.~(\ref{nup}) suggests that this
case is acceptable provided that
\beq
m_\sigma\gsim 220~{\rm GeV}.
\label{msigma}
\eeq
The superpotential for the SM singlet
superfields is comprised only of the terms
[cf. Eq.~(\ref{W})]
\beq
W_{\rm singlet}=[\lambda_0S^8+
\lambda_1S^4(PQ)^2+\lambda_2(PQ)^4]/
M_{\rm P}^5.
\eeq
The resulting scalar potential is shown in
Fig.~\ref{fig}.

\par
Using Eq.~(\ref{epszeta}), we obtain the value
of the amplification factor necessary for the
model to work:
\beq
\varepsilon\sim 10^{-8.5}\,\Omega_{\rm dec}.
\label{e5W}
\eeq
Now, let us assume, at first, that
\mbox{$\Omega_{\rm dec}<1$}, i.e. the curvaton
decays before domination. Then,
Eq.~(\ref{epszeta1}) gives
\beq
\varepsilon\sim 10^{-4.75}
g^{\frac{1}{2}}\Omega_{\rm dec}^{\frac{1}{2}},
\eeq
which means that
\beq
g\sim 10^{-7.5}\,\Omega_{\rm dec}.
\eeq
This value lies well within the range given in
Eq.~(\ref{grange}).

\par
With these values, it is straightforward to
show that all the relevant constraints are
satisfied. For example, the constraint in
Eq.~(\ref{H*}), in the case when
\mbox{$\Omega_{\rm dec}<1$}, becomes
\begin{equation}
g\,\Omega_{\rm dec}\sim(\varepsilon\zeta)^2
\left(\frac{M_{\rm P}}{m_{3/2}}\right)
^{\frac{n+2}{n+1}},
\end{equation}
which, for \mbox{$n=5$} and the value of $g$
found above, can be easily checked to hold
true. According to Eq.~(\ref{Treh}), the
universe reheats at temperature
\beq
T_{\rm reh}\sim 10^3\,\Omega_{\rm dec}
~{\rm GeV},
\eeq
which is far from challenging the gravitino
bound.

\par
However, the PQ symmetry breaking scale is
found from Eq.~(\ref{fPQ}) to be
\beq
v_0\sim 10^{15.5}~{\rm GeV}.
\label{v00}
\eeq
Such a high PQ scale results in axion
overproduction unless the original axion
misalignment angle $\theta_{\rm a}$ is much
less than unity. Indeed, in this case,
Eq.~(\ref{thaxbound}) suggests that
\beq
\theta_{\rm a}\leq 10^{-3}
\eeq
should be enough to avoid axion overproduction.
The above constraint is much less stringent
than the constraint on the present value of
$\theta_{\rm a}$ from $CP$ violation in
strong interactions (coming from experimental
bounds on the electric dipole moment of the
neutron): \mbox{$\theta_{\rm a}<10^{-9}$}, but
it still undermines the motivation for the PQ
symmetry, which is meant to explain how we can
get the present value of $\theta_{\rm a}$ so
small starting with a natural value of the
initial $\theta_{\rm a}\sim 1$. This is why it
is preferable to consider the case when the
curvaton decays {\em after} domination, where
the axion energy density can be efficiently
diluted by the entropy production from the
curvaton decay.

\par
When the curvaton decays after domination, we
have \mbox{$\Omega_{\rm dec}\approx 1$}. Then,
Eq.~(\ref{e5W}) becomes
\beq
\varepsilon\sim 10^{-8.5}.
\label{e-value}
\eeq
Also, from Eq.~(\ref{gW2}), we obtain the bound
\beq
g>10^{-7.5}.
\label{g-value}
\eeq
As a result of the above, the reheat
temperature after the end of inflation is
found, from Eq.~(\ref{Treh}), to be
\beq
T_{\rm reh}\sim g\sqrt{m_{3/2}M_{\rm P}}>10^3~
{\rm GeV}.
\eeq
Satisfying the gravitino bound
(\mbox{$T_{\rm reh}\leq 10^9~{\rm GeV}$}) sets
a weak upper bound on $g$: \mbox{$g\leq 0.03$}.
However, this bound
is relaxed by the dilution of the gravitinos
due to the entropy production
by the curvaton decay. Indeed, in this case,
according to Eq.~(\ref{Treh1}), the
hot big bang begins at the temperature
\beq
T_{\rm REH}\sim 10~{\rm MeV},
\label{TREH}
\eeq
which is close to the BBN bound, but does not
violate it.

\par
Now, from Eq.~(\ref{SS}), we find the
entropy production ratio to be
\beq
\frac{{\cal S}_{\rm after}}
{{\cal S}_{\rm before}}\sim
10^{7.5}g.
\eeq
The relic abundance of axions can be calculated
by applying the formulae of Ref.~\cite{turner},
where we take the QCD scale
$\Lambda_{\rm QCD}=200~{\rm MeV}$ and ignore
the uncertainties for simplicity. Comparing
this relic abundance with the best-fit value of
the cold dark matter (CDM) abundance in the
universe from the measurements of WMAP
\cite{index}, we find that, for $v_0$ as large
as the one shown in Eq.~(\ref{v00}), the
entropy of the universe must be increased by a
factor of order $10^5$ at least. In the
present model, this is actually an overestimate
since, as it turns out, the axions are
generated, in the viable cases, after the
domination of the oscillating curvaton field.
Axion production in a universe dominated by a
coherently oscillating scalar field has been
studied in Ref.~\cite{lazaetc}. In such a
universe, axion oscillations begin at a smaller
temperature than in a radiation dominated
universe. Consequently, the initial ratio of
the axion number density to the photon number
density is smaller. This then implies that, in
this case, less entropy production is required.
Taking this effect carefully into account, we
obtain the bound
\beq
g>10^{-4.5},
\label{glast}
\eeq
which is more stringent than the bound in
Eq.~(\ref{g-value}), as expected. A large
coupling between the inflaton field and its
decay products can be realized if the VEV of
the $s$ modulus corresponds to a point of
enhanced symmetry. Note that, in this case,
there is no moduli problem because $s$ decays
much earlier than BBN.

\par
Moreover, we have to make sure that the entropy
production occurs after the onset of the axion
oscillations. The latter, according to
Eq.~(\ref{Ta}), occurs at energy density given
by
\beq
\rho^{\frac{1}{4}}_{\rm axosc}\sim 1~{\rm GeV},
\label{Taxosc}
\eeq
where we used Eq.~(\ref{v00}). Comparing the
above equation with Eq.~(\ref{TREH}), we see
that the requirement in Eq.~(\ref{axbound}) is
satisfied. Thus, we conclude that axion
overproduction can be avoided, due to entropy
release at curvaton decay.

\par
It should be noted, however, that this
mechanism of diluting the axions by the entropy
produced when the curvaton decays after
dominating the universe may lead \cite{choi} to
a cosmological disaster. Generally, a sizable
fraction of the curvaton's decay products
consists of sparticles, which eventually turn
into stable lightest sparticles (LSPs) in
models (such as ours) with an unbroken matter
parity symmetry. The freeze-out temperature of
the LSPs is typically much higher than the
reheat temperature in Eq.~(\ref{TREH}).
Actually, it is even higher than the energy
density scale corresponding to the onset of
axion oscillations (see Eq.~(\ref{Taxosc})).
Consequently, the LSPs freeze out immediately
after their production and can, subsequently,
overclose the universe leading to a
cosmological catastrophe.

\par
In the present model, the PNGB curvaton can
decay into a pair of squarks, sleptons,
charginos or neutralinos with a decay width
which can be comparable to the width of its
main decay channel to a pair of Higgs particles
(see Eq.~(\ref{Gs})). We should, however,
observe that our curvaton, being a SM singlet,
couples to the charginos and neutralinos only
via their Higgsino component. Taking $\mu$ to
be much greater than the soft masses of the
bino and wino, which is often encountered in
various realizations of MSSM, we can ensure
that the lighter charginos and neutralinos are
predominantly gauginos. We can further choose
all the squark, slepton and heavier chargino
and neutralino masses to exceed half the
curvaton mass (see Eq.~(\ref{msigma})) so that
all the curvaton decay channels involving these
particles are kinematically blocked. Thus, the
only decay channels allowed are to a pair of
lighter charginos or neutralinos. The
corresponding rates can be easily suppressed
by reducing the Higgsino component of these
sparticles. Indeed, assuming that their
Higgsino components are about $1\%$, we obtain
a suppression factor of order $10^{-8}$. In
view of the fact that these sparticles have
comparable masses to the Higgs particles, we
then conclude that only a fraction of about
$10^{-8}$ of the energy density of the universe
soon after the curvaton decay consists of LSPs.
This fraction is enhanced by a factor of about
$10^7$ until the time $t_{\rm eq}$ of equal
matter and radiation energy densities when the
cosmic temperature
$T_{\rm eq}\sim 10^{-9}~{\rm GeV}$
and remains essentially constant thereafter.
So, the cosmological disaster from the possible
overproduction of LSPs at the curvaton decay
can be avoided. Moreover, the LSPs can
contribute to the CDM in the universe.

\par
Using Eq.~(\ref{epsmin}) with \mbox{$n=5$}, we
find
\beq
\varepsilon_{\rm min}\sim 10^{-12.5}.
\eeq
Therefore, Eqs.~(\ref{See}) and (\ref{e-value})
suggest that
\beq
|S|_*\sim 10^4H_*.
\label{cond}
\eeq
The above can, in principle, be used in
Eqs.~(\ref{SN}) and (\ref{SNbound})
to constrain the parameters
of the underlying model (e.g. $F_s$).

\par
A useful quantity to calculate in order to
evaluate Eqs.~(\ref{SN}) and (\ref{SNbound}) is
the number of e-foldings which corresponds to
the cosmological scales $N_*$. The cosmological
scales range from a few times the size of the
horizon today $\sim H_0^{-1}$ down to scales
$\sim 10^{-6}H_0^{-1}$ corresponding to masses
of order $10^6M_\odot$ with $M_\odot$ being
the solar mass. Typically, this spans about 13
e-foldings of inflation. For the estimate of
$N_*$, we will choose a scale roughly in the
middle of this range. More precisely, we will
take the scale that re-enters the horizon at
the time when structure formation begins, i.e.
at the time $t_{\rm eq}$ of equal matter and
radiation energy densities. Then it is
straightforward to obtain
\beq
\frac{\exp(N_*)}{H_*^{\frac{1}{3}}
t_{\rm eq}^{\frac{1}{2}}}\sim
\left(\frac{\Gamma_\sigma\Gamma_{\rm inf}}
{H_{\rm dom}}\right)^{\frac{1}{6}}\sim
\frac{(m_\sigma^3M_{\rm P}^4)^{\frac{1}{6}}}
{v_0},
\eeq
where we have used Eqs.~(\ref{Gs}) and
(\ref{Hdom}) with \mbox{$\sigma_{\rm osc}\sim
v_0$}. Note that, remarkably, the above is
independent from $g$. Putting \mbox{$H_*\sim
m_\sigma\sim m_{3/2}$}, we obtain
\beq
N_*\simeq 38,
\label{N*}
\eeq
where we have also taken Eq.~(\ref{v00}) into
account. The number of e-foldings that
corresponds to the decoupling of matter and
radiation (when the CMBR is emitted) is roughly
\mbox{$N_*+1.5$}, while the one which
corresponds to the present horizon is about
\mbox{$N_*+9$}.

\par
Using the above, let us attempt to investigate
first the case when
\beq
2F_sN_{\rm x}\ll 1.
\label{2FN}
\eeq
Substituting Eq.~(\ref{SNo}) in
Eq.~(\ref{NNN}), we obtain
\beq
\frac{\left(22.1-\ln\frac{N_{\rm x}}{N_*}
\right)\left(\frac{N_{\rm x}}{N_*}-1\right)}
{76\left[\frac{N_{\rm x}}{N_*}\left(\ln
\frac{N_{\rm x}}{N_*}-1\right)+1\right]}\ll 1,
\label{NNNll}
\eeq
where we also used Eqs.~(\ref{SN0}),
(\ref{cond}) and (\ref{N*}), and the fact that
$H^2\simeq H_{\rm m}^2(1-e^{-2F_sN})$, which,
in this case, is $\approx 2F_sNH_{\rm m}^2$. In
order to find how small the left hand side
(LHS) of this inequality should actually be, we
must observe that the contribution to the
spectral index of density perturbations $n_s$
originating from the evolution of $v$ during
inflation is $-2H_*^{-1}(\dot{v}/v)_*$, which
is negative. Moreover, one can easily check
that, in the present example, all the other
contributions \cite{iso} to the spectral tilt
for the curvaton are negligible. This is due to
the fact that $\varepsilon\ll 1$ (see
Eq.~(\ref{e-value})) and, as it turns out, also
$c\ll 1$ (see below). Using Eq.~(\ref{fdotf}),
one can further show that, under these
circumstances, $dn_s/d\ln k$ is very small and,
thus, no running of the spectral index is
predicted in our model. For fixed $n_s$,
the recent results of WMAP imply \cite{index}
that $n_s=0.96\pm 0.02$. Therefore, at $95\%$
c.l., $n_s\geq 0.92$. It is then obvious that the
LHS of the inequality in Eq.~(\ref{NNNll})
should not exceed about $0.04$. This
requirement is met provided that
\beq
\frac{N_{\rm x}}{N_*}\gsim 500,
\eeq
which, in view of Eq.~(\ref{N*}), implies that
\beq
N_{\rm x}\gsim 1.9\times 10^4.
\label{N0}
\eeq
From Eq.~(\ref{SNo}), we then obtain that
\beq
c_S\lsim 2.4\times 10^{-4}.
\label{cS1}
\eeq
It can be checked that, with these values, the
requirement in Eq.~(\ref{SNbound0}) is well
satisfied. From Eqs.~(\ref{2FN}) and (\ref{N0}),
one obtains
\beq
F_s\ll 2.63\times 10^{-5}\quad\Rightarrow\quad
c\ll 7.89\times 10^{-5},
\label{c}
\eeq
where we also used Eq.~(\ref{Fs}). Such a small
$c$ implies that modular inflation is not
really of the fast-roll type and may last for a
large number of e-foldings. Indeed, according
to Eq.~(\ref{Ntot}),
\mbox{$N_{\rm tot}\gg 10^6$}. In view of
Eqs.~(\ref{Fs}) and (\ref{Hx}),
Eqs.~(\ref{cS1}) and (\ref{c}) suggest that
\beq
|m_S|\lsim 1.55\times 10^{-2}H_*\quad
{\rm and}\quad m_s\ll 8.88\times 10^{-3}H_*
\label{R}
\eeq
with $H_*\lsim 4.47\times 10^{-2}H_{\rm m}$.
The above values for $c$ and $c_S$ are
plausible (requiring only mildly tuned masses)
but not very pleasing, since we would prefer
\mbox{$c\sim c_S\sim 1$}. (Note that larger
values of $N_{\rm x}$ result in more severe
tuning of $c$ and $c_S$.) The small values
obtained may be due to the condition in
Eq.~(\ref{2FN}), which we imposed to simplify
the problem.

\par
Therefore, let us consider, now, that
\beq
2F_sN_*\ll 1\quad{\rm and}\quad
2F_sN_{\rm x}\gg 1.
\eeq
In this limit, Eq.~(\ref{nnn}) takes the form
\beq
-\frac{22.1+\ln (2F_sN_*)}{76\ln (2F_sN_*)}
\ll 1.
\label{nnnllgg}
\eeq
We find that the LHS of this inequality remains
smaller than $0.04$ provided that
$2F_sN_*\lsim 0.004$, which yields
\beq
F_s\lsim 5.26\times 10^{-5}
\quad\Rightarrow\quad
c\lsim 1.58\times 10^{-4}.
\label{Fsllgg}
\eeq
This is less fine-tuned than the values in
Eq.~(\ref{c}). In order to estimate the lower
bound on $N_{\rm x}$ corresponding to the upper
limit on $F_s$, we approximate Eq.~(\ref{nnn})
for $2F_sN_*\ll 1$, but any value of
$2F_sN_{\rm x}$:
\beq
\frac{[22.1-\ln(\frac{1-e^{-2F_sN_{\rm x}}}
{2F_sN_*})](1-e^{-2F_sN_{\rm x}})}
{76[(1-e^{-2F_sN_{\rm x}})\ln(
\frac{e^{2F_sN_{\rm x}}-1}{2F_sN_*})
-2F_sN_{\rm x}]}\ll 1.
\eeq
For $2F_sN_*\simeq 0.004$, the LHS of this
inequality is kept smaller than $0.04$ even if
$2F_sN_{\rm x}$ becomes as low as 4, which
yields
\beq
N_{\rm x}\gsim 3.8\times 10^4.
\label{Nxllgg}
\eeq
Saturating the bounds in Eqs.~(\ref{Fsllgg})
and (\ref{Nxllgg}), i.e. taking
\beq
F_s\simeq 5.26\times 10^{-5}~(c\simeq 1.58
\times 10^{-4})\quad {\rm and}
\quad N_{\rm x}\simeq 3.8\times 10^4,
\label{bestchoice}
\eeq
we find from Eq.~(\ref{SN}) that
\beq
c_S\simeq 4.92\times 10^{-4},
\label{cS2}
\eeq
which is a little more natural than the values
in Eq.~(\ref{cS1}). It can be checked that,
with the values in Eq.~(\ref{bestchoice}), the
requirement in Eq.~(\ref{SNbound}) is well
satisfied. Using Eq.~(\ref{Ntot}), it is easy
to see that, in this case, the total number of
e-foldings of inflation is
\beq
N_{\rm tot}\simeq 6.6\times 10^5.
\eeq
In view of Eqs.~(\ref{Fs}) and (\ref{Hx}),
Eq.~(\ref{bestchoice}) suggests that
\beq
|m_S|\simeq 2.22\times 10^{-2}H_*\quad{\rm and}
\quad m_s\simeq 1.26\times 10^{-2}H_*
\label{bestmasses}
\eeq
with $H_*\simeq 6.32\times 10^{-2}H_{\rm m}$,
which are not very different from the upper
bounds in Eq.~(\ref{R}). After some
investigation, it can be realized that not much
improvement can be made on the results.

\par
In conclusion, we see that, in the $n=5$ case
with $\theta\sim 1$, our model can work,
typically, for values
\beq
c_S,~c\lsim{\cal O}(10^{-4})
\eeq
or, equivalently, for masses
\beq
|m_S|,~m_s\lsim {\cal O}(10^{-2})\,H_*,
\eeq
where \mbox{$H_*\sim m_{3/2}\sim 1~{\rm TeV}$}.
Such values are quite natural and imply only a
mild tuning on the masses of the rolling field
$|S|$ and the inflaton modulus. This is
necessary because the effective mass
$|\bar m_S|$ should
remain small enough during the relevant part of
inflation for $|S|$ to be slowly rolling and
the constraint in Eq.~(\ref{nsbound}) to be
satisfied. The condition for this is that $c_S$
or, equivalently, $|m_S|$ be small
\cite{footnote8}. Yet, a substantial variation
of $|S|$ from the phase transition until the
time when the cosmological scales exit the
horizon is necessary for obtaining the required
value of the amplification factor
$\varepsilon^{-1}$ (recall
that the required $\varepsilon$ is always much
bigger than $\varepsilon_{\rm min}$). This is
achieved with a small mass for the inflaton
modulus, which leads to a large number of
e-foldings. So, modular inflation cannot be of
the fast-roll type in this case. The above
findings are similar to the ones in
Ref.~\cite{lett}, despite the fact that there
the example studied considered the case when
the curvaton decayed before domination with
\mbox{$n=2$}. This suggests that the above
results are quite robust.

\par
One may wonder why, since both the inflaton
$s$ and the field $|S|$ turn out to be
light when the cosmological scales exit the
inflationary horizon, we cannot use those
fields to generate the observed curvature
perturbation. The reason is that, in contrast
to the PNGB curvaton, the perturbations of
those fields are not amplified. Hence, their
contribution to the overall curvature
perturbation is insignificant. Indeed, for the
inflaton, we have
\beq
\zeta_s\simeq\frac{1}{5\sqrt 3\pi}
\frac{V_*^{\frac{3}{2}}}{|V'|_*M_{\rm P}^3}
\simeq\frac{3}{5\pi}\left(\frac{H_*}{m_s}
\right)^3\frac{m_s}{M_{\rm P}}\,e^{F_sN_*},
\label{zetas}
\eeq
where the prime denotes derivative with respect
to the inflaton $s$ and we have used
Eqs.~(\ref{Vinf}) and (\ref{Fs}). For the above
discussed values in Eqs.~(\ref{N*}),
(\ref{bestchoice}) and (\ref{bestmasses}) and
for $H_*\sim m_{3/2}\sim 1~{\rm TeV}$,
Eq.~(\ref{zetas}) gives \mbox{$\zeta_s\sim
10^{-12}$}, which is much smaller than the
observed value \mbox{$\zeta\simeq 2\times
10^{-5}$}. Similarly, for $|S|$, we have
\beq
\zeta_{|S|}\simeq
\frac{2}{3}\left.\frac{\delta|\bar S|}
{|\bar S|}\right|_*
\sim\frac{H_*}{v_0}\sim\varepsilon\zeta_\sigma,
\eeq
where
\mbox{$|\bar S|\equiv\big||S|-|S|_0\big|$},
$\delta|\bar S|$ the perturbation in $|\bar S|$
and we have used the fact that
\mbox{$|S|_0\sim v_0\gg |S|_*$}
with $|S|_*$ given by Eq.~(\ref{cond}). For the
values discussed above,
\mbox{$\zeta_{|S|}\sim 10^{-13}\ll\zeta$}, where
we considered that
\mbox{$\zeta\approx\zeta_\sigma$}.

\section{Discussion and conclusions}
\label{sec:concl}

\par
In this paper, we have studied modular
inflation, which uses a string axion as the
inflaton field. The inflationary scale
(\mbox{$\sim 10^{10.5}~{\rm GeV}$}) is
determined by the scale of gravity mediated
soft SUSY breaking. Such low-scale inflation,
even though it can still solve the flatness and
horizon problems of the standard hot big bang
cosmology, cannot generate the observed
curvature perturbation (necessary to explain
the CMBR anisotropy and structure formation)
from the quantum fluctuations of the inflaton
field. However, we have shown that this type of
modular inflation {\em can} generate the
appropriate amplitude of superhorizon curvature
perturbation to account for the observations
through the use of a suitable curvaton field.

\par
The curvaton field that we have used is a PNGB
which is an angular degree of freedom
orthogonal to the QCD axion field (that we
called the orthogonal axion) in a class of
SUSY PQ models. We considered models that
generate the PQ scale dynamically (by using
flaton fields), while they also solve the $\mu$
problem of MSSM. In these models, one needs more
than one SM singlet superfields to break the
global ${\rm U}(1)_{\rm PQ}$ symmetry (with the
exception of its matter parity subgroup).
Hence, apart from the axion, there is at least
one other angular degree of freedom (the
orthogonal axion) which may be kept
appropriately light during inflation and can be
responsible for the curvature perturbation in
the universe. This could be achieved if the
potential possesses a valley of minima with a
negative inclination and the system happens to
slowly roll down this valley during the
relevant part of inflation with some of the
SM singlet fields acquiring values much smaller
than their vacuum values. Under these
circumstances, the orthogonal axion may be kept
light during inflation and its perturbation
from inflation may be later amplified as the SM
singlets acquire their vacuum values accounting
for the observed curvature perturbation in the
universe.

\par
Following this promising idea, we have
attempted to construct appropriate curvaton
models using two SM singlet superfields $P$ and
$Q$, charged under the PQ symmetry. However, we
have shown that it is not possible to construct
suitable curvaton models by using only two SM
singlet superfields, because, in this case, the
orthogonal axion mode cannot avoid being
massive during the relevant part of inflation.
Hence, we studied PQ models which
involve a third superfield $S$ and specified
the general conditions under which these models
can contain a suitable PNGB curvaton. Actually,
we have shown that they possess a shifted
valley of minima at almost constant values of
$|P|$ and $|Q|$ of the order of their vacuum
values. The value of $|S|$, however, which
parametrizes the valley, can be kept much
smaller than its vacuum value during the
relevant part of inflation. Also, there exists
an orthogonal axion which remains light during
inflation and can serve as curvaton.

\par
For definiteness, we considered a concrete
class of models of this category where the
superfield $S$ has vanishing PQ charge (see
also Ref.~\cite{chun}). These models are simple
extensions of MSSM based on the SM gauge group
and possessing, besides the global anomalous PQ
symmetry, a global ${\rm U}(1)$ R symmetry and
a discrete $Z_2$ symmetry. The superpotential
of the models includes all the terms satisfying
these symmetries. The baryon and lepton numbers
are automatically conserved to all orders in
perturbation theory as a consequence of the R
(and the PQ) symmetry.

\par
To study the cosmology, we have also taken into
account that the SM singlet fields in these
models ($P$, $Q$ and $S$) are expected to
receive SUGRA corrections to their soft
SUSY-breaking masses-squared (as well as their
soft SUSY-breaking trilinear $A$-terms) of
order set by the Hubble parameter. We consider
the corrections to the masses-squared to be
positive in all cases. Since our inflation
model has $H_*$ of order the electroweak scale,
which is also the scale of the soft masses,
these SUGRA corrections do not seriously affect
the physics with the exception of the $S$
field, whose soft mass-squared is taken to be
negative. This is intentional in order to
facilitate a phase transition during inflation
which sends $|S|$ rolling away from the origin
along the shifted valley and reaching its
true minimum by the end of inflation (cf.
Ref.~\cite{lett}). As a result, the order
parameter of our PNGB curvaton field increases
substantially after the cosmological scales
exit the inflationary horizon. This amplifies
the curvaton perturbation according to the
mechanism presented in Ref.~\cite{amplif} and
the observed value of the curvature
perturbation in the universe can be achieved
despite the low inflation energy scale.

\par
We have investigated in detail the above
scenario and showed that the requirements for a
successful curvaton put important constraints
both on the choice of model and also on the
model parameters. Indeed, we have shown that
only a few members of our class of PQ
models are eligible for successful curvaton.
We then have concentrated on a particular such
model and, using natural values for the model
parameters, we have studied analytically its
performance as curvaton model. We found that
the model can indeed work successfully in the
context of modular inflation with only a mild
tuning of the inflaton's mass and the mass of
$S$: \mbox{$|m_S|,m_s\lsim 0.01H_*$}. The
bound on $|m_S|$ comes from the requirement
that $|S|$ be slowly rolling when the
cosmological scales exit the horizon,
otherwise the scale invariance of the spectrum
of curvature perturbations will be
destabilized. The bound on $m_s$, on the other
hand, originates from the large variation of
$|S|$ between the phase transition and the exit
of the cosmological scales from the horizon,
which is necessary for obtaining the correct
amplification of the curvaton perturbation.
This bound implies that modular inflation
cannot be of the fast-roll type and may last
for a large number of e-foldings.

\par
In this model, the PQ scale turns out to be
quite large (it is comparable to the scale of
grand unification). It is actually well above
the standard cosmological bound from the
requirement that the universe is not overclosed
by axion overproduction. However, overclosure
of the universe can be avoided, in this model,
by diluting the primordial axions through the
entropy release by the decay of the curvaton
field. For this to be effective, the inflaton
modulus has to decay early enough so that the
curvaton may well dominate the radiation
background. Consequently, we need a
comparatively large decay coupling constant for
the inflaton, which is possible if the VEV of
the inflaton corresponds to an enhanced
symmetry point.

\par
After the curvaton decays, the hot big bang
begins. This occurs not long before BBN. So,
baryogenesis has to take place soon after the
time of the curvaton decay at the latest.
Moreover, it is clear that, in the present case
where the curvaton decays after dominating the
energy density of the universe, baryons must be
generated through the decay of the curvaton (or
of some of its decay products), since otherwise
there will be \cite{luw} an unacceptably large
baryon isocurvature perturbation. It is also
obvious that leptogenesis \cite{lepto} (or
electroweak baryogenesis) cannot work here
since the reheat temperature is very low for
the non-perturbative electroweak sphaleron
effects to operate. Therefore, almost the only
viable option is that the observed baryon
asymmetry of the universe is directly generated
by the decay of the curvaton (or of its decay
products). One possibility is that the curvaton
has suitable baryon (and lepton) number
violating decay channels via non-renormalizable
Lagrangian operators of higher order with decay
widths comparable to its main decay width to
Higgs particles. This, of course, requires an
appropriate extension of our model, which, as
it stands, has exact baryon (and lepton) number
conservation. It may be possible \cite{benakli}
to obtain the required Lagrangian operators by
embedding our model in a larger scheme with
(large) extra dimensions. This baryogenesis
issue deserves further study, which we postpone
for the future.

\par
The curvaton model analyzed in this paper can
be considered to accommodate \cite{juan}
low-scale inflationary models other than
modular inflation of the type considered here.
However, modular inflation due to string axions
is one of the better theoretically motivated
low-scale inflation models in the literature.

\section*{ACKNOWLEDGEMENTS}
\par
We would like to thank D.H. Lyth and
J. McDonald for discussions. This work was
supported by the European Union under the
contract MRTN-CT-2004-503369.

\def\ijmp#1#2#3{{Int. Jour. Mod. Phys.}
{\bf #1},~#3~(#2)}
\def\plb#1#2#3{{Phys. Lett. B }{\bf #1},~#3~(#2)}
\def\zpc#1#2#3{{Z. Phys. C }{\bf #1},~#3~(#2)}
\def\prl#1#2#3{{Phys. Rev. Lett.}
{\bf #1},~#3~(#2)}
\def\rmp#1#2#3{{Rev. Mod. Phys.}
{\bf #1},~#3~(#2)}
\def\prep#1#2#3{{Phys. Rep. }{\bf #1},~#3~(#2)}
\def\prd#1#2#3{{Phys. Rev. D }{\bf #1},~#3~(#2)}
\def\npb#1#2#3{{Nucl. Phys. }{\bf B#1},~#3~(#2)}
\def\npps#1#2#3{{Nucl. Phys. B (Proc. Sup.)}
{\bf #1},~#3~(#2)}
\def\mpl#1#2#3{{Mod. Phys. Lett.}
{\bf #1},~#3~(#2)}
\def\arnps#1#2#3{{Annu. Rev. Nucl. Part. Sci.}
{\bf #1},~#3~(#2)}
\def\sjnp#1#2#3{{Sov. J. Nucl. Phys.}
{\bf #1},~#3~(#2)}
\def\jetp#1#2#3{{JETP Lett. }{\bf #1},~#3~(#2)}
\def\app#1#2#3{{Acta Phys. Polon.}
{\bf #1},~#3~(#2)}
\def\rnc#1#2#3{{Riv. Nuovo Cim.}
{\bf #1},~#3~(#2)}
\def\ap#1#2#3{{Ann. Phys. }{\bf #1},~#3~(#2)}
\def\ptp#1#2#3{{Prog. Theor. Phys.}
{\bf #1},~#3~(#2)}
\def\apjl#1#2#3{{Astrophys. J. Lett.}
{\bf #1},~#3~(#2)}
\def\n#1#2#3{{Nature }{\bf #1},~#3~(#2)}
\def\apj#1#2#3{{Astrophys. J.}
{\bf #1},~#3~(#2)}
\def\anj#1#2#3{{Astron. J. }{\bf #1},~#3~(#2)}
\def\apjs#1#2#3{{Astrophys. J. Suppl.}
{\bf #1},~#3~(#2)}
\def\mnras#1#2#3{{MNRAS }{\bf #1},~#3~(#2)}
\def\grg#1#2#3{{Gen. Rel. Grav.}
{\bf #1},~#3~(#2)}
\def\s#1#2#3{{Science }{\bf #1},~#3~(#2)}
\def\baas#1#2#3{{Bull. Am. Astron. Soc.}
{\bf #1},~#3~(#2)}
\def\ibid#1#2#3{{\it ibid. }{\bf #1},~#3~(#2)}
\def\cpc#1#2#3{{Comput. Phys. Commun.}
{\bf #1},~#3~(#2)}
\def\astp#1#2#3{{Astropart. Phys.}
{\bf #1},~#3~(#2)}
\def\epjc#1#2#3{{Eur. Phys. J. C}
{\bf #1},~#3~(#2)}
\def\nima#1#2#3{{Nucl. Instrum. Meth. A}
{\bf #1},~#3~(#2)}
\def\jhep#1#2#3{{J. High Energy Phys.}
{\bf #1},~#3~(#2)}
\def\lnp#1#2#3{{Lect. Notes Phys.}
{\bf #1},~#3~(#2)}
\def\appb#1#2#3{{Acta Phys. Polon. B}
{\bf #1},~#3~(#2)}
\def\njp#1#2#3{{New J. Phys.}
{\bf #1},~#3~(#2)}
\def\pl#1#2#3{{Phys. Lett. }{\bf #1B},~#3~(#2)}
\def\jcap#1#2#3{{J. Cosmol. Astropart. Phys.}
{\bf #1},~#3~(#2)}
\def\mpla#1#2#3{{Mod. Phys. Lett. A}
{\bf #1},~#3~(#2)}


\begin{thebibliography}{99}
\bibitem{footnote1}
In 11-dimentional SUGRA, the axions correspond
\cite{Cfield} to the massless modes of the
third-rank antisymmetric tensor
$C=b_{\mu\nu}dx^{11}\wedge
dx^\mu\wedge dx^\nu+b_I\partial_{_{11}}
\omega^I_{\alpha\beta^*}dx^{11}\wedge dy^\alpha
\wedge dy^{\beta *}$, where
$x^{11}$ is the coordinate of the so-called
``eleventh-segment'', i.e. the one extra
dimension which is somewhat larger than the
others. 
\bibitem{footnote3}
This requirement may be even more fundamental
in origin. Indeed, a PNGB with rapidly varying
order parameter cannot be treated as an
effectively free field. We would like to thank
D.H. Lyth for pointing this out.
\bibitem{footnote4}
Note that the effective mass term of $S$ which
should replace the mass term of $S$ in
Eqs.~(\ref{V1}) and (\ref{V1angles}) as the
system rolls down the shifted valley is
much smaller in magnitude, but this has no
effect on our previous discussion other than
to reduce drastically the slope of this
valley.
\bibitem{footnote5}
As mentioned, in order for the local minimum
given by $x_{+}$ in Eq.~(\ref{xpm}) to be a
global minimum, we need \mbox{$|\bar{A}|>
|\bar{A}|_{\rm min}\equiv(n+3)m$}. Enforcing
this constraint, the lower bound on $Z$
becomes \mbox{$Z>(n+1)/(n+3)$}. Hence, it is
easy to show that, for the entire range of
$|\bar{A}|$ (\mbox{$|\bar{A}|_{\rm min}<
|\bar{A}|<+\infty$}), the value of $\kappa$
changes only by a factor of
$(3/4)[(n+3)/(n+2)]^2$.
\bibitem{footnote6}
For cosmic temperatures which are much higher
than
\mbox{$\Lambda_{\rm QCD}\simeq 200~{\rm MeV}$},
the zero-temperature mass of the axion
$m_{\rm a}$ must be replaced by the
finite-temperature mass: \mbox{$m_{\rm a}(T)
\sim m_{\rm a}(\Lambda_{\rm QCD}/T)^4$}
\cite{turner,gpy}. However, in our case, it
turns out that the relevant temperatures are
never too high for temperature corrections to
be vital. This is due to the large values of
$f_{\rm a}$ encountered here. So, the use of
the zero-temperature axion mass is adequate
for our rough order-of-magnitude estimates.
Moreover, in the interesting case where the
onset of axion oscillations takes place during
curvaton domination (see Sec.~\ref{sec:ex}),
the temperatures involved are even smaller and
temperature corrections to the axion mass do
not come into effect.
\bibitem{footnote7}
Saturating the bound in Eq.~(\ref{gW2}) is
equivalent to curvaton decay at the moment of
domination, in which case
\mbox{${\cal S}_{\rm after}/
{\cal S}_{\rm before}\sim 1$}.
\bibitem{footnote8}
This has no effect on our previous discussion
other than to reduce the slope of the shifted
valley.
\end{thebibliography}
\end{document}